\documentclass[pra,superscriptaddress ,twocolumn, aps,fleqn,floatfix]{revtex4-1}

\usepackage{amsmath}
\usepackage{amstext}
\usepackage{amsfonts}
\usepackage{graphicx}
\usepackage{draftcopy}
\usepackage{flafter}
\usepackage[below]{placeins}
\usepackage{float}
\usepackage{tabularx}
\usepackage{times}

\def\ket#1{\left|#1\right\rangle}
\def\bra#1{\left\langle#1\right|}
\def\avg#1{\left\langle#1\right\rangle}

\begin{document}

\title{A Couped-Qubit Tavis Cummings Scheme for Prolonging Quantum Coherence}

\author{Amrit De}
\affiliation{Department of Physics and Astronomy, University of California - Riverside, CA 92521}

\date{\today}

\begin{abstract}
Qubit-qubit interactions can significantly boost quantum coherence times for Bell states. The coherence-time-enhancements are however not monotonic and there exists a phase where further increasing the interaction is unhelpful. A resonator in a suggested circuit QED type implementation of the Tavis-Cummings(Dicke) model, is shown to shift this transition point depending on the number of loaded photons. This allows the resonator to amplify the coherence enhancements in certain regimes. The interactions also induce unusual collapse and revival type behavior for the entanglement dynamics.
A new and exact open quantum systems formalism -- the quasi-Hamiltonians for the Dicke model thus reveals how a Bell state in a resonator can be protected against $1/f$ noise from randomly fluctuating two level systems.  Simple circuit level details are given for flux qubits.
\end{abstract}

\maketitle

\section{Introduction}\label{sec.intro}
Qubits are typically formed from anharmonic systems such as impurities, quantum dots (QD) or Josephson junction (JJ) devices that have unevenly spaced energy levels. In order to `lock-in' to the two eigenstates that constitute the qubit, they are resonantly coupled to a system with evenly spaced energies -- a harmonic oscillator. Of the many such possible cavity-QED systems\cite{Raimond2001,Leibfried2003,Hennessy2007}, superconducting (SC) qubits coupled to transmission line resonators or LC oscillators\cite{Devoret1989,Schoelkopf2008,Martinis2009} have drawn a tremendous amount of attention as it allows for extremely strong resonator-qubit coupling\cite{Niemczyk2010}. This easily facilitates interconversion between solid-state- and photonic qubits, it allows the distant transmission of quantum information and the resonators can be used for qubit read out.


However these circuit-QED type setups are not closed quantum systems, which results in the loss of quantum coherence. Of their many possible decoherence mechanisms, such as non-cavity mode decay, vacuum fluctuations and electrical noise\cite{Blais2004} -- $1/f$ noise is one of the hardest to deal with and is arguably the leading cause of decoherence for most SC qubits\cite{Clarke2008}.

Flux qubits, Phase qubits and the Quantronium are all susceptible to $1/f$ flux noise\cite{Yoshihara2006,Sank2012,Ithier2005}. Charge qubits such as Cooper-pair-boxes(CPB) are typically susceptible to $1/f$ charge noise\cite{McDermott2009,Paladino2014}. Though CPB variants like the Transmon, split-Transmon and Xmons are largely insensitive to charge noise, they not necessarily completely immune to flux noise for various reasons\cite{Koch2007pra,Paladino2014}.

Though $1/f$ flux noise was first observed in SQUIDs quite some time ago\cite{Koch1983,Wellstood1987}, its exact microscopic origins are not fully understood. It is thought that fluctuating two-level-systems(TLSs) at the at the JJ's metal-insulator interface\cite{Yoshihara2006,Bialczak2007} lead to flux noise. Due to the recent interest in quantum computing, a number of models have attempted to explain this phenomenon\cite{Koch2007,deSousa2007,Faoro2008,Choi2009,Chen2010,Kechedzhi2011,De2014arxiv,Atalaya2014}. Phenomenologically the TLSs behave like Ising spins that randomly flip in time. Experimental evidence also suggests the presence of long range magnetic order among the TLSs\cite{Sendelbach2008,Bluhm2009,Anton2013}

\begin{figure}
\centering
\includegraphics[width=0.85\columnwidth]{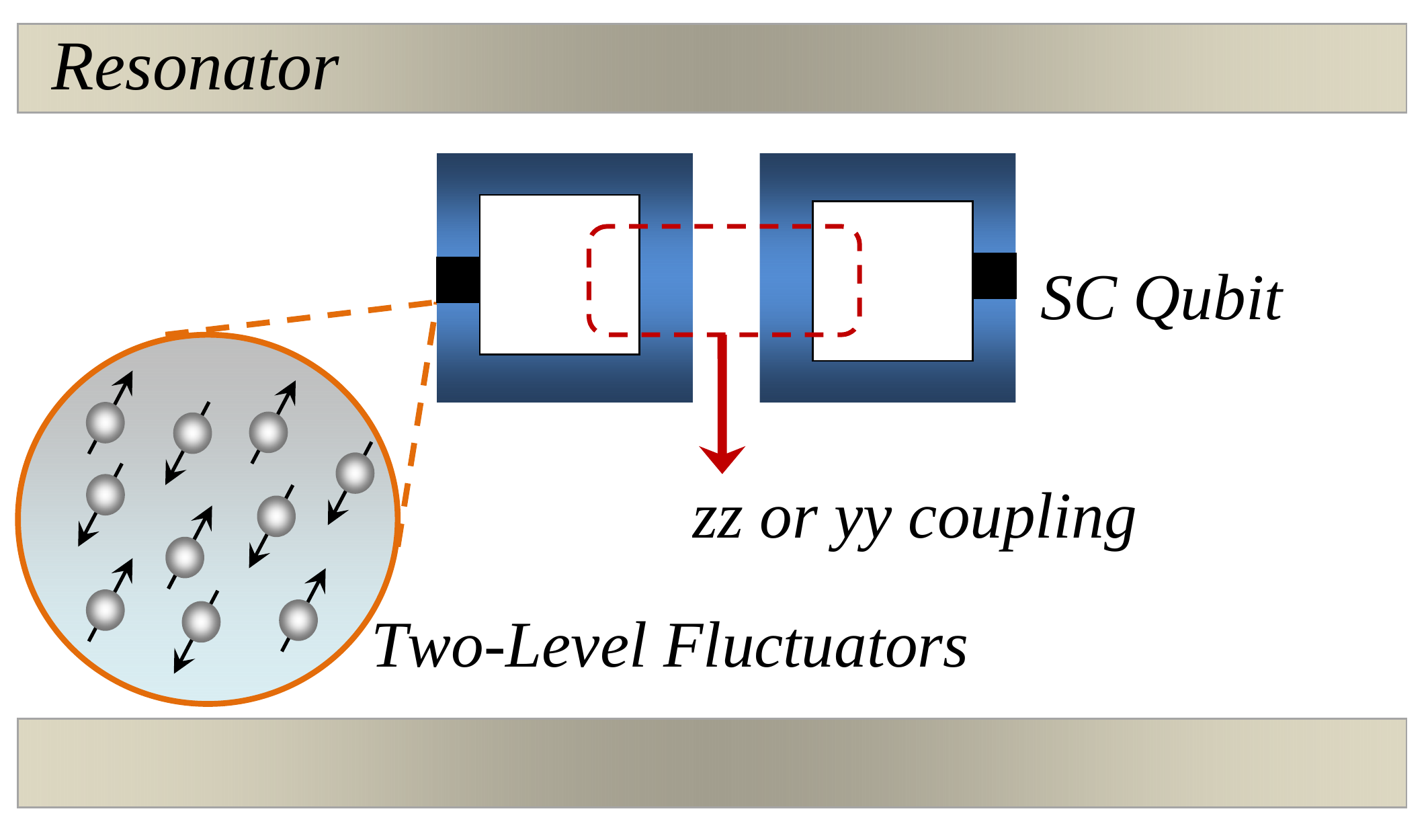}
\caption{(Color online) Two superconducting qubits with $\sigma_z\sigma_z$ or $\sigma_y\sigma_y$ interactions forming a Bell state and in a resonator cavity. A strongly entangled Bell state is less suseptible to local $1/f$ noise which arises from the several fluctuating two-level systems, present at the SQUID's metal-insulator interface.}
\label{fig:sch}
\end{figure}

One way to minimize the effects of the TLSs is by suitably engineering the Josephson tunnel junction\cite{Martinis2005,Oh2006}. Other ways, that can be used in conjunction, would be to use techniques such as dynamical decoupling(DD)\cite{Khodjasteh2005,Uhrig2007}, topological protection\cite{Kitaev:2001} and decoherence free subspaces(DFS)\cite{Bacon2000,Lidar2008}. Active methods, such quantum error correcting codes(QECC)\cite{Calderbank1996,Steane1996} can also be used for additional protection in case of a bit-flip or phase-flip error (or any linear combination). However QECCs require a lot of physical qubits, frequent monitoring, and moreover the error rate per qubit must be small for the QECC to be effective\cite{Knill-error-bound2}. Hence the qubits need to be inherently less susceptible to noise in the first place.

In general, $1/f$ noise is a problem for other types of qubits as well such as those based on semiconductor QDs\cite{Vandersypen2004,MacLean2007,Taubert2008,Kuhlmann2013,Paladino2014}, where random telegraphic noise arises due to the electrons tunneling between the reservoir and the QD. As cavity QED quantum information processing schemes involving QDs typically rely on strong coupling to a photonic crystal's cavity mode\cite{Imamoglu1999,Hennessy2007}, the implementation suggested in this paper will also be applicable to these semiconductor systems.

Resonators operating in the dispersive regime are usually used for qubit read out\cite{Blais2004,Cooper2004,Siddiqi2006} where loading more photons allows for greater discernability. A different role is suggested here for the resonator and the loaded photons. In this paper a passive technique is proposed to protect non-local logical qubits against local $1/f$ noise in a suggested circuit-QED(or cavity-QED) type setup. More specifically, qubit-qubit interactions($J$) are shown to significantly boost the coherence times of an entangled Bell state qubit in a resonator if the noise on one qubit is minimally correlated to the other. The coherence enhancements are however non-monotonic and there is a phase where further increasing $J$ is detrimental. The resonator can shift this transition point and the shift depends on the average photon number. In some regimes for large $J$, the resonator cavity can in fact amplify the coherence-enhancements as more photons are loaded into the system.

If the noise on each qubit is uncorrelated then the interaction induced suppression of decoherence for the Bell State qubits is somewhat like DD. DD works by applying pulsed unitary-rotations to flip the qubit's precession about an axis orthogonal to the noise. In the Bell State basis however such uncorrelated dephasing noise will cause pole to pole oscillations on the effective Bloch sphere\cite{De2011} and the $J$ now acts orthogonally to this and internally generates precession which counters the noise. This interaction based technique can be thought of as a rotated typical DD Hamiltonian. Also, much like DD this method is most effective for longer noise correlation times\cite{De2011}. In general, for multi-qubit networks, it is possible to implement universal gates sets and manipulate entangled states on bipartite lattices, where the interactions are always turned on, by using scalable pulse sequences\cite{De2013prl,De2014pra}.

Since a key requirement for this method to work is that the noise on one qubit be minimally correlated with the noise on the other qubit, having two physically separate but interacting JJ device qubits forming the logical Bell state qubit is well suited for this. In addition having the two qubits inside the resonator induces new photon number dependent phases for the coherence improvements. Multi-Qubit Tavis-Cummings models(TCMs) have been experimentally realized using different types of SC qubits\cite{Majer2007,Fink2009,Chen2014}. Also in general, two qubit gate operations have been realized for semiconductor based singlet-triplet Bell state qubits\cite{Shulman2012}.


A very recently introduced quasi-Hamiltonian formalism\cite{Cheng2008,Joynt2011} is key in uncovering these results. Typically Lindblad master equations(LME) are used to model dissipation\cite{Carmichael.book}. However the LMEs are phenomenological and don't provide a connection to the underlaying microscopic noise mechanisms. Moreover since the LME inherently assumes that the noise is Markovian, the decoherence effects becomes irreversible and hence LMEs are not particularly useful for modeling the restoration of quantum coherence via unitary operations such as DD or the method suggested here.

In the quasi-Hamiltonian($\mathcal{H}_q$) method, the cavity-qubit dynamics is combined with TLS's stochastic dynamics in a manner similar to path integrals, so that the ensemble averaged dynamics of the entire system can be obtained from the single-shot non-unitary evolution of a non-Hermitian Hamiltonian. Following an earlier derivation for the Jaynes-Cummings model\cite{De2013pra}, the quasi-Hamiltonian for the TCM in the presence of $N$ TLSs is presented in this paper.

The model and the quasi-Hamiltonian method are presented in the next section. In Sec.\ref{sec.res}, the main results are discussed, followed by a flux qubit example with details at the circuit level in Sec\ref{sec.cir}. Some comments are made on the role of LMEs in the context of this paper in Sec.\ref{sec.Lind}.



\section{The Model and Method}\label{sec.model}
Consider two SC qubits coupled to a single quantized mode of a resonator as shown in fig.\ref{fig:sch}. At the metal insulator interface of the qubits there exist a number of fluctuating TLS which flip randomly in time. The system Hamiltonian is: $H = {H}^{sys} + {H}^{fluc}$ where, the Tavis-Cummings term with qubit-qubit interactions is
\begin{eqnarray}
{H}^{sys}&=&\omega_r{a^{\dagger}a} + J_{\nu\nu}{\sigma}_\nu^{(1)}{\sigma}_\nu^{(2)} \label{Hjc2Q} \\\nonumber
&&+\frac{1}{2}\sum_{k=1,2}\omega_o^{(k)}\sigma_z^{(k)} + 2\lambda_k\left(a\sigma_+^{(k)} + a^\dagger\sigma_-^{(k)} \right),
\end{eqnarray}
and the noise term from the fluctuating TLSs is
\begin{eqnarray}
{H}^{fluc}=\displaystyle\sum_{j,k}s_j^{(k)}(t)\left[{{\bf g}_j^{(k)}\cdot{\boldsymbol\sigma}^{(k)}} +{\zeta}_j^{(k)}\left(a\sigma_+^{(k)} + a^\dagger\sigma_-^{(k)}\right)\right]
\label{Hn}
\end{eqnarray}
The TCM (or Dicke model) is block diagonal in the $\{\ket{00;n+1},\ket{01;n},\ket{10;n},\ket{11;n-1}\}$ subspace as only the energy conserving terms are retained in the rotating wave approximation. Here, $\omega_o^{(k)}/2$ is the energy separation between the qubit's excited state $|1\rangle$ and the ground state $|0\rangle$, $\omega_r$ is the resonator frequency, $a^\dagger(a)$ is the photon creation(annihilation) operator for a single cavity mode. Let $\Delta_k=\omega_r-\omega_o^{(k)}$ be the field detuning.


The first term in $H^{fluc}$ results from the fluctuating TLSs at the metal insulator interface which are modeled as randomly flipping Ising spins. Here $s_j^{(k)}(t)$ is the $j^{th}$ TLS, coupled to qubit-$k$ (where $k\in\{1,2\}$), that switches randomly between $\pm1$ and ${\bf g}_j^{(k)}=[g_{xj}^{(k)},g_{yj}^{(k)},g_{zj}^{(k)}]$ is the corresponding noise amplitude vector. If the TLSs are statistically independent, the noise on the qubit is just the sum of all contributions from individual TLSs, whose correlation function is:
\begin{equation}
\langle{s_i(t_1)s_j(t_2)}\rangle\propto\exp(-2\gamma_j|t_1-t_2|)\delta_{ij}.
\end{equation}
Here $\gamma_j$ is the TLS's switching rate. The last term in Eq.\ref{Hn} is due to fluctuations in the dipole coupling caused by the resonator. The randomly fluctuating TLSs cause energy fluctuations in the qubit (depending on the noise- and qubit-basis) which perturbs the wave function $|1\rangle\rightarrow|1\rangle'$ and could cause fluctuations in $\lambda\propto\langle{0}|\hat{d}|{1}\rangle'$, inducing an off-diagonal energy relaxation term in the qubit subspace\cite{De2013pra}.

For the temporal dynamics, assume that the qubits and the single resonator mode are initially unentangled pure states, {\it i.e.} $\rho(0)=\rho_q(0)\otimes\rho_r(0)$, where the SC qubits initially form a Bell state $\rho_q(0)=|\psi_{+}\rangle\langle\psi_+|$ (where $\ket{\psi_\pm}=\ket{01}\pm\ket{10}/\sqrt{2}$) and the resonator mode is $\rho_r(0)={\sum_n}|C_n|^2|n\rangle\langle n|$. The overall time dependent density matrix evolves as $\rho(dt)=U[\rho_q(0)\otimes\rho_r(0)]U^{\dagger}$ where $U=\exp(-iH dt)$. By taking a partial trace over the cavity mode, the reduced 2 qubit density operator, $\rho'_q(dt) = Tr_r[\rho(dt)]$, can be expressed in terms of the 16 component {\it generalized} Bloch vector ${\boldsymbol\eta}(t)$:
\begin{equation}
\rho'(dt)= \frac{1}{4}\displaystyle\sum_{\mu,\nu}\eta_{\mu\nu}(dt)\left[\sigma _{\mu}\otimes \sigma _{\nu}\right]
\label{rho-SU4}
\end{equation}
where $\{\mu,\nu\}\in\{0,x,y,z\}$, $\sigma _{0}=\mathbf{I}$, $\eta_{00}=1$, the other $\eta_{\mu\nu}$ are the 15 components of the Bloch vector and $\sigma_{\mu}\otimes\sigma_{\nu}=\Lambda'_{\mu\nu}=\Lambda _{m}$ (except $\mu=\nu=0$) are $SU(4)$ generators. Noting that $\rho'(dt)=U\rho'(0)U^{\dagger}$ and using the identity $Tr\left[\Lambda _{m}\Lambda _{l}\right] =4\delta _{ml}$, Eq.\ref{rho-SU4} can be transformed into a transfer matrix equation which governs the qubit dynamics:
\begin{equation}
\boldsymbol{\eta}(dt)=\boldsymbol{\mathcal T}\boldsymbol{\eta}(0)
\label{eq.eta}
\end{equation}
where $\mathcal{T}_{ml}=\sum_n|C_n|^2Tr\left[ U\Lambda_{m}U^{\dagger }\Lambda _{l}\right]/2$ are the matrix elements of $\boldsymbol{\mathcal{T}}$.

While the random stochastic dynamics of the TLSs is governed by the master equation \cite{VanKampen.book}:
\begin{equation}
\frac{d\mathbf{W}(t)}{dt}=\mathbf{VW{\rm\it(t)}}\label{eq.WV}
\end{equation}
where $\mathbf{V}$ and $\mathbf{W}$ are respectively the transition-rate-matrix and flipping-probability-matrix. And so $\mathbf{W}=\exp(-\mathbf{V}t)$.
By combining the stochastic and quantum transfer matrices in a path integral type manner, in the small time limit, for all possible TLS configurations and noting that this equates to ${\small\displaystyle\lim_{dt\rightarrow{0}}}\exp(-i\mathcal{H}_qdt)$ \cite{Cheng2008,Joynt2011,Zhou2010,De2013pra} one obtains the time-independent non-Hermitian quasi Hamiltonian that allows the \emph{exact} single shot calculation of the ensemble averaged temporal dynamics of the TCM with the fluctuators: $\hat{\mathcal{H}}^{(n)}_q = \hat{\mathcal{H}}_{q}^{(sys)} + \hat{\mathcal{H}}_{q}^{(fluc)}$
{\small
\begin{eqnarray}
&&\hat{\mathcal{H}}^{(sys)}_{q} = i{\mathcal I}\otimes \Big[ \displaystyle\sum_{\nu=x,y,z}J_{\nu\nu}(L'_\nu\otimes\Sigma_{\nu}+\Sigma_{\nu}\otimes{L'_\nu})+\\\nonumber
&&(\Delta_1{L'_z}-\lambda'_1\sqrt{n+1}{L'_x})\otimes{L'_0}+ L'_0\otimes{(\Delta_2{L'_z}-\lambda'_2\sqrt{n+1}{L'_x})}\Big],
\label{Hq_tc}
\end{eqnarray}}
{\small
\begin{eqnarray}
&~&\hat{\mathcal{H}}^{(fluc)}_{q} = i\displaystyle\sum_{j=1}^N\gamma_j(\sigma_x^{(j)}-{\mathcal I})\otimes{L'_0}\otimes{L'_0} + \\\nonumber
&~& i\left[\displaystyle\sum_{j=1}^{M}\sigma_z^{(j)}\otimes\mathbf{g}_j\cdot\mathbf{L'}\otimes{L'_0}
+ \displaystyle\sum_{j=M+1}^{N}\sigma_z^{(j)}\otimes{L'_0}\otimes\mathbf{g}_j\cdot\mathbf{L'}\right].  \\\nonumber
\label{Hq_n}
\end{eqnarray}}
Here $M$ uncorrelated TLSs coupled to one qubit and $N-M$ to the other, $L'_{i=x,y,z}$ are the $4\times4$ form of the $SO(3)$-generators\cite{De2011}, $L'_0$ is the identity, $\mathbf{L'}=[L'_x,L'_y,L'_z]$, ${\mathcal I}$ is a $2^N$ dimensional identity,
$2\Sigma_{x}=\Lambda'_{ox}+\Lambda'_{zx}$ ,
$2\Sigma_{y}=\Lambda'_{xo}+\Lambda'_{xz}$ and
$2\Sigma_{z}=\Lambda'_{xx}-\Lambda'_{yy}$.
The final ensemble averaged time dependent $16$ component 2-qubit Bloch vector is obtained from the following projection:
\begin{equation}\small
{\boldsymbol\eta}(t)=\langle f_N|...\otimes\langle f_1|\sum_n|C_n|^2e^{-i{\mathcal H}^{(n)}_{q}t}|i_1\rangle\otimes...|i_N\rangle{\boldsymbol\eta}(0)
\label{THqn}
\end{equation}
where $|i_j\rangle $ and $|f_j\rangle$ are the initial and final state vectors of the $j^{th}$ TLS, that satisfy $\mathbf{W}_j|i_j(f_j)\rangle =|i_j(f_j)\rangle $. For an unbiased TLS: $|i_j\rangle =|f_j\rangle =[1,1]/\sqrt{2}$. Note that $\mathcal{\hat H}^{(sys)}_q$ by itself returns an exact solution for just the TCM model. In general the two qubits and the resonator form a tripartite system and there are four non-equivalent ways to partition this into bipartite subsystems\cite{Tessier2003}. For the entanglement dynamics of the Bell states, the marginal density operator is considered with the cavity mode traced out to calculate the concurrence -- a measure of two-body entanglement\cite{Wootters1998}:
\begin{equation}
{\mathcal{C}}(\rho)=max\left[\sqrt\kappa_1-\sqrt\kappa_2-\sqrt\kappa_3-\sqrt\kappa_4,0\right]
\label{Con}
\end{equation}
where, $\kappa _{m=1,2,3,4}$ are the eigenvalues of $\rho(\sigma _{y}\otimes \sigma _{y})\rho ^{\ast }(\sigma _{y}\otimes \sigma _{y})$ in decreasing order. $\rho(t)$ is obtained from Eqs.\ref{rho-SU4} and \ref{THqn}.

\section{Results and Discussion}\label{sec.res}
\begin{figure}
\includegraphics[width=1\columnwidth]{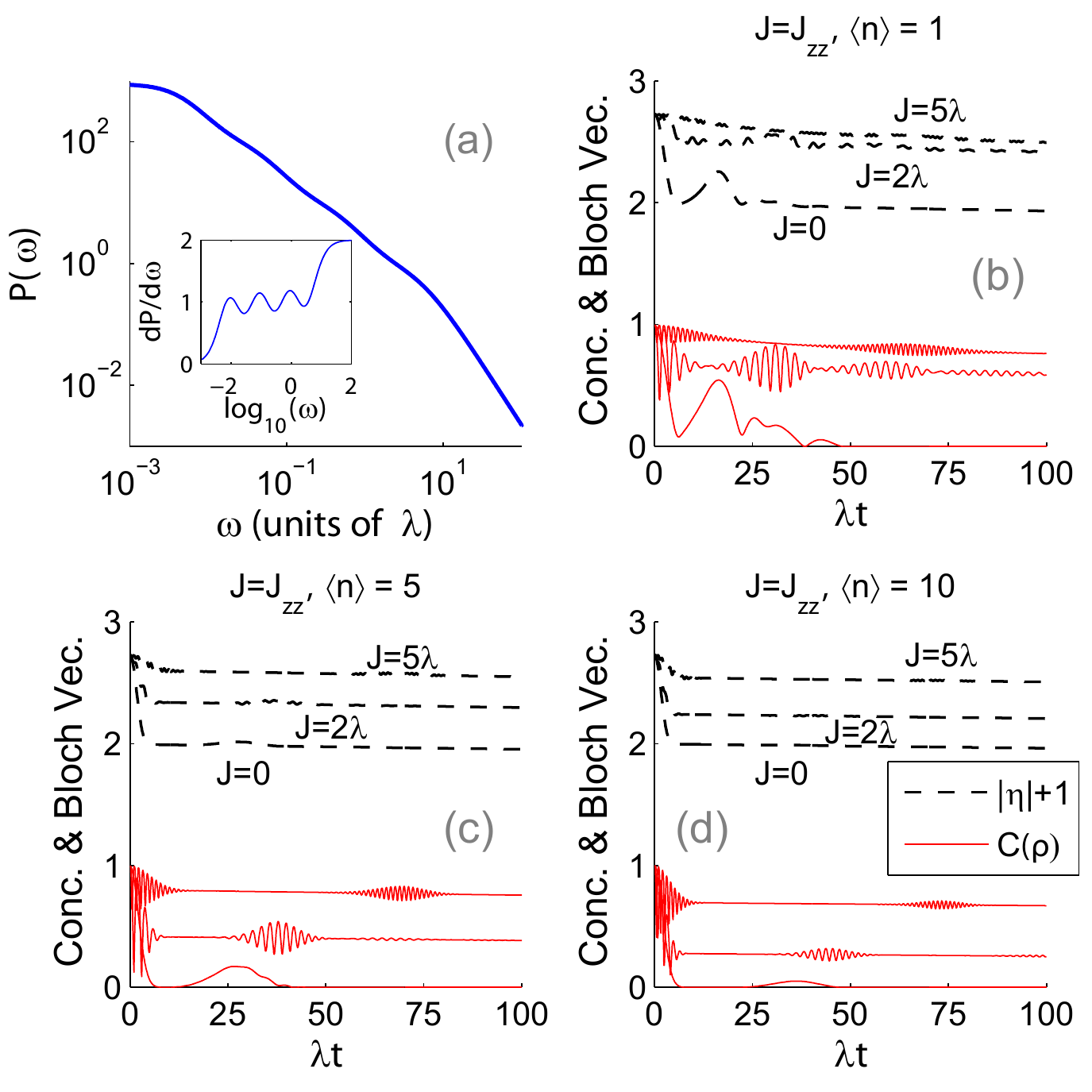}
\caption{\small(color online) {\bf(a)} Noise spectrum of 4 TLSs that each qubit is subjected to. The $1/f$ noise is indicated by the slope $dP/d{\omega}\sim 1$ in the inset.
Magnitude of the Bloch vector, $\eta$ (offset by $+1$) and the concurrence, $\mathcal{C}(\rho)$, for the $\psi _{\pm}=\ket{01}\pm\ket{10}/\sqrt{2}$ Bell states shown as a function of normalized time ($\lambda{t}$) and $J_{zz}$ coupling in the presence of $1/f$ noise for {\bf(b)} $\avg{n}=1$ {\bf(c)} $\avg{n}=5$ and {\bf(d)} $\avg{n}=10$ photons. Note that $\mathcal{C}$ has the same dependence on $J$ as $\eta$.}
\label{CBV1}
\end{figure}

A $1/\gamma$ distribution of $\gamma$s is taken to generate a $1/f$ noise power spectrum, $P(\omega)=\sum_j\int\avg{s_j(0)s_j(t)}e^{-i\omega t}dt$ as shown in Figs. \ref{CBV1}-(a). An individual qubit's lifetime is calculated with $\avg{n}=0$ and $\langle\sigma_{z(x)}(t)\rangle$ is fit to $\exp(-t/T_{2(1)})$ at short times. The noise parameters are then adjusted so that for a typical value of $\lambda=100$ MHz, the resulting $T_{1(2)}$ is very similar to experiment\cite{Bylander2011}. Here the fitted noise parameters are $g_z^j\approx0.2\lambda,~g_x^j\approx g_y^j=3g_z^j$ so that the individual qubit's lifetimes with $\avg{n}=0$ are $T_1=400\lambda^{-1}$ and $T_2=100\lambda^{-1}$. Note that $g_z=20$ MHz is comparable to the TLS splitting energies reported so far (the average is about $10$ MHz \cite{Yoshihara2006,Zhou2012} while the largest is $45$ MHz \cite{Yoni2010}).

\begin{figure}
\centering
\includegraphics[width=1\columnwidth]{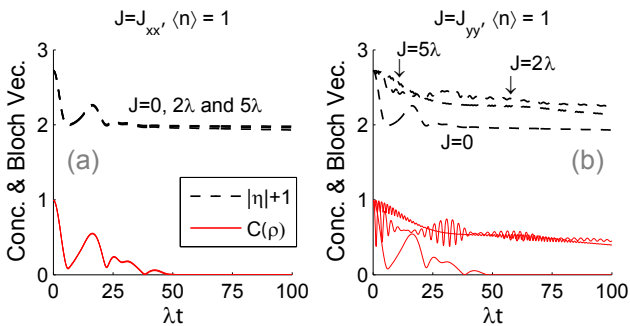}
\caption{\small(color online) Same as fig.\ref{CBV1} but for {\bf(a)} $J_{xx}$ and {\bf(b)} $J_{yy}$ qubit-qubit interactions with $\langle{n}\rangle=1$ photon.}
\label{CBV2}
\end{figure}

In Fig. \ref{CBV1}(b), the magnitude of the Bloch vector,$\left\vert{\boldsymbol\eta}\right\vert$ (a measure of purity), and the concurrence,$\mathcal{C}(\rho)$, is shown as a function of time for various interaction strengths $J_{zz}$ for the qubits initially prepared in the $\psi _{+}$ Bell state and the single resonator mode initially in a coherent state $|C_n|^2=\exp(-\langle n \rangle){{\langle n \rangle}^{n}}{/n!}$. The average number of photons, $\langle{n}\rangle$, are increased from fig. \ref{CBV1}(b)-(d). These calculations are done by exactly solving Eq.\ref{THqn} numerically. In Fig. \ref{CBV2}, the dissipative dynamics is shown for $J_{xx}$ and $J_{yy}$ interactions for $\langle{n}\rangle=1$. It is clear that while $J_{zz}$ strongly suppresses the decoherence, $J_{yy}$ also works but is not as effective and almost no improvements are seen for $J_{xx}$. While $\vert\boldsymbol\eta\vert$ and $\mathcal{C}(\rho)$ tend to track each other they are not in one-to-one correspondence. Furthermore the concurrence displays collapse and revival type behavior which depends on $J$ and is infact more distinct for small $\langle{n}\rangle$.

\begin{figure}
\centering
\includegraphics[width=1\columnwidth]{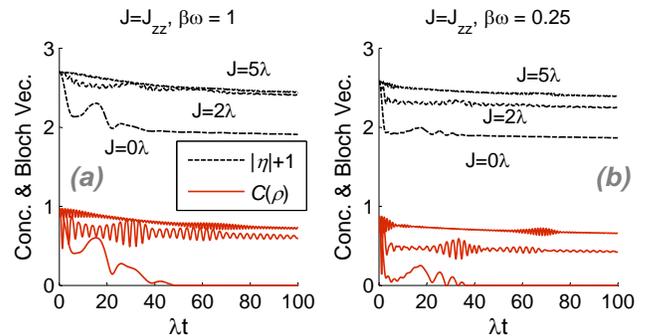}
\caption{\small(color online) Bloch vector ($\eta$) and concurrence ($\mathcal{C}$) for the Bell states with $J_{zz}$ interactions in the presence of $1/f$ noise and with the resonator initially in a \emph{thermal state} with {\bf(a)} $\beta\omega=1$ corresponding to $\langle{n}\rangle=0.58$ and {\bf(b)} $\beta\omega=0.25$ or $\langle{n}\rangle=3.52$. $\mathcal{C}$ and $\eta$ have the same dependence on $J$.}
\label{thermal}
\end{figure}

Note that though a coherent state was chosen for the resonator, it makes no qualitative difference for these effects if some other state, for e.g., a thermal state is chosen as shown in Fig.\ref{thermal}, where $|C_n|^2 = \langle{n}\rangle^n/(\langle{n}\rangle+1)^{1+n}$ and $\langle{n}\rangle=[\exp(\beta\omega_r)-1]^{-1}$.

To understand how $J$ suppresses decoherence consider an effective Bloch sphere with $\psi_\pm$ poles in the presence of pure dephasing noise\cite{De2011}. In the absence of any coupling, the dephasing noise causes the ensemble averaged effective Bloch vector to dissipatively oscillate between the $\psi _{+}$ and $\psi _{-}$ poles through the center, since averaging restores chiral symmetry. Now, when $J$ is turned on in a direction orthogonal to the poles, the effective Bloch vector will tend to precess about one of the poles -- which essentially counters the dissipation due to the noise, which is trying to drive $\eta$ towards the center of the Bloch sphere.

The above effective single qubit picture is not applicable in the presence of energy relaxation and the resonator -- this needs the full $15$ component Bloch vector. For a better understanding $\mathcal{H}_q$ is expanded to $4^{th}$ order in time. For an initial Bell state,
the non-zero Bloch vector components are $\eta_{ox}=\eta_{xo}$, $\eta_{xx}$, $\eta_{yy}$, $\eta_{yz}=\eta_{zy}$, $\eta_{zz}$. Also note that the resonator breaks the symmetry between $x$ and $y$. Consider $\eta_{yy}$, as it is most representative of the behavior shown in Figs.\ref{CBV1} and \ref{CBV2}:
{\small
\begin{eqnarray}
\eta_{yy}(t)\approx \sum_n |C_n|^2 \bigg( 1 - (\lambda_n^2+\frac{1}{2}\sum_j{\bf g}^2_j)t^2 + \frac{1}{3}\sum_j\gamma_j{\bf g}^2_j t^3  ~~~~\\\nonumber
 +\Big[ \lambda_n^2(J_{zz}^2 + \lambda_n^2) + (J_{xx}^2+ 3\lambda_n^2)\sum_j{g_{zj}^{2}} + (J_{zz}^2 + 6\lambda_n^2 )\sum_j{g_{xj}^{2}}\\\nonumber
  +\sum_j{\bf g}^2_j(1-4\gamma_j^2) + 6{\bf g}^2_1{\bf g}^2_2 \Big]\frac{t^4}{24} \bigg)
\end{eqnarray}\label{nyy1}
}
where ${\bf g}_j^2=g_{xj}^2+g_{zj}^2$ and $\lambda_n=\lambda\sqrt{n+1}$.

The effect of the interaction is seen earliest in $t^{4}$. Though $J_{zz}$ is key in suppressing decoherence, its effectiveness also depends on $\lambda$ as apparent from the $J_{zz}^2\lambda_n^2$ term. Both $J_{zz}$ and $J_{xx}$ couple to noise components orthogonal to them, but this term is ineffective as the $g$s are small. {\emph Note} that the main reason why $J_{xx}$ does not affect the dynamics (see Fig.\ref{CBV2}.(b)) is because this interaction term simply commutes with the $\lambda_n(\sigma_-^{(j)}a^\dagger+\sigma_+^{(j)} a)$ term that is also along $x$. From this it follows that a non-commuting $J_{yy}$ term will also suppress the decoherence (as shown in Fig.\ref{CBV2}.(a)). The introduction of just $J_{yy}$ leads to new terms $-\lambda_n g_{yj}g_{zj}t^3$ and $J^2_{yy}(\lambda_n^2+\sum{\bf g}^2_j)t^4$ in Eq.\ref{nyy1} (also set $J_{xx}=J_{zz}=0$). $J_{yy}$'s effectiveness in preserving coherence is however affected by its coupling to $z$ and $x$ noise components. If both $J_{yy}$ and $J_{zz}$ are introduced then this leads to cross terms like $-J_{zz}J_{yy}\lambda^2t^4$, which are sign dependent but could be useful in some cases\cite{Geller2014}.

Apart from the TLSs, the resonator also decoheres the qubits as more photons, $\langle{n}\rangle$, are loaded as shown in Figs.\ref{CBV1}(b)-(c). However the minimum floor that $\eta$ (and $\mathcal{C}$) can decohere to, is set by $J$. This is better understood by considering pure dephasing noise in the weak noise coupling limit so that $[\mathcal{H}_q^{sys},\mathcal{H}_q^{fluc}]\approx0$), which gives:
{\small
\begin{eqnarray}
\eta_{yy}(t) \approx\prod_j{\sum_{n}} |C_n|^2\frac{\lambda_n^2\cos(\varphi t)+J^2}{\lambda_n^2+J^2}\zeta_j(t) \label{nyy2} \\
\zeta_j(t)=\left[\cos(\Omega{t})+\frac{\gamma_j}{\Omega_j}\sin(\Omega_j t)\right]e^{-\gamma_j t} \label{zeta}
\end{eqnarray}}
where $\varphi=\sqrt{J^2+\lambda_n^2}$, $\Omega_j=\sqrt{g_z^2-\gamma_j^2}$ and $J=J_{zz}$ or $J_{yy}$. Note that all the figures shows exact numerical calculations.

In the absence of $J$ and the noise, each $\cos(\lambda\sqrt{n+1}t)$ term in the summation represents Rabi oscillations (weighted by $|C_n|^2$). At $t=0$ these terms are perfectly correlated, however at longer times destructive and then constructive interference between these terms leads to the collapse and revival of Rabi oscillations (CRR). Generally CRR is more distinct when $\avg{n}$ increases. A very intriguing feature here is that for higher $J$, $\eta_{yy}$ (or $\mathcal{C}(\rho)$ in figs.\ref{CBV1} and \ref{CBV2}) show very distinct and persistent CRR even just for $\langle{n}\rangle=1$. This is because as $J$ increases, the minimum floor of the collapse and the frequency $\varphi$ (associated with different photon excitations) also increases.

The dissipation term, Eq.\ref{zeta}, can be classified into non-Markovian ($g_z>\gamma $), Markovian ($g_z<\gamma $) and an intermediate regime ($g_z\sim\gamma$). For Markovian noise, the trigonometric functions become hyperbolic functions leading to a monotonically decaying the Bloch vector. For $1/f$ noise, the $g_j$s fall in between the selected range of $\gamma_j$s resulting in a mixture. The Markovian noise part mostly tends to dominate and washout the oscillatory behavior of the non-Markovian noise, though small oscillations superposed on top of a smoothly decaying function are visible in fig.\ref{CBV1}-(b). Once the $\avg{n}$ is increased the resonator induced dissipation will completely dominate as seen in fig.\ref{CBV1}-(c) and (d). Much like DD, the method proposed in this paper works best when the noise is non-Markovian or the noise correlation times are large.


\begin{figure}
\includegraphics[width=1\columnwidth]{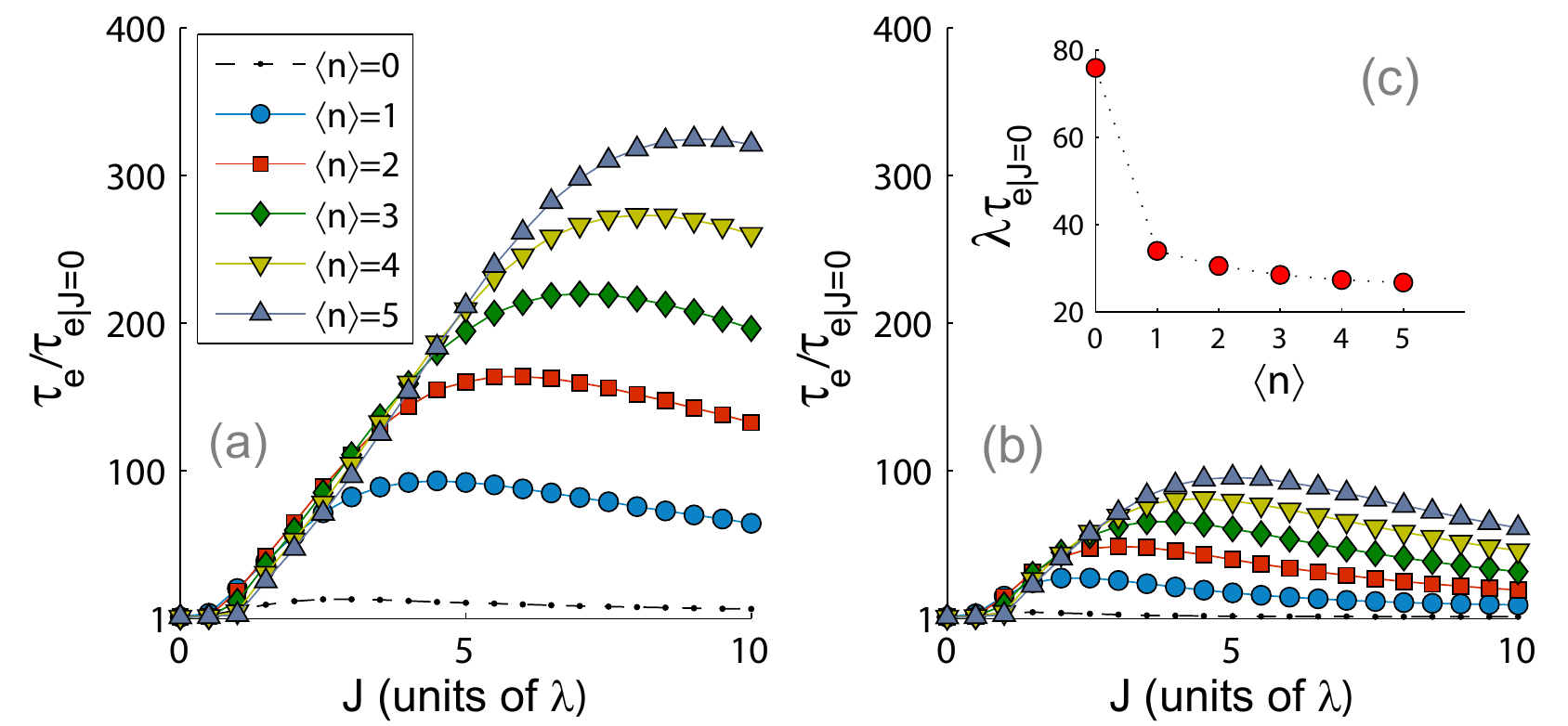}
\caption{\small(color online) Improvements in the normalized envelope function coherence time, $\tau_{e}/(\tau_e|J=0)$, for the $\psi_\pm$ qubit as a function of the coupling strength $J$ and average photon number $\langle{n}\rangle$ for (a) $J=J_{zz}$ and (b) $J=J_{yy}$. (c) $\tau_{e}$ as a function of $\avg{n}$ for $J=0$. Enhancements with $J$ for $\tau_{e}$ saturate unless further amplified by loading more $\avg{n}$. Note that the $1/f$ noise parameters are the same as in the other figures.}
\label{TvsJ}
\end{figure}

In general, for two entangled qubits the dissipative behavior cannot be characterized just by $T_1$ and $T_2$. However, in order to understand how the overall decay will scale with $J$, the envelope of ${\boldsymbol\eta}(t)$  is fit to $\exp({-t}/{\tau_{e}})$ from which the decay constant $\tau_{e}$ is extracted. In fig.\ref{TvsJ}, $\tau_e$ (normalized to $\tau_e|J=0$) is shown as a function of $J_{zz}$ and $J_{yy}$ for various $\avg{n}$. It is seen that initially $\tau_{e}$ increases rapidly as a nonlinear function of $J$, which is a very encouraging since even small improvements in the maximum-achievable-$J$ can lead to significantly more robust Bell state qubits.

However, when integrated over longer times, $\tau_e$ does not monotonically keep increasing with $J$. There exists a phase where $\tau_e$ begins to decrease with increasing $J$ beyond a certain threshold. Though the initial $d\tau_e/dJ$ slope is always greater for smaller $\avg{n}$, the eventual threshold for the rollover ({\it i.e.} $d\tau_e/dJ=0$) increases with $\avg{n}$. These resonator mediated pseudo amplification features can be explained from the more exact analytical form for $\eta_{yy}$ (which is very ungainly). However this expression for $\eta_{yy}$ has factors of the form:
$|C_n|^2\cos\left[t\left( J^2 + \lambda_n^2 + g_z^2 \pm\sqrt{ (J^2-g_z^2)^2 + 2J^2\lambda_n^2} \right)^{\frac{1}{2}}\right]$.
In one case, the argument tends to zero for large $J$ leading to an unhelpful slowly decaying cosine function in the first quadrant.
For a fixed $\lambda_n$, a large $J$ will lead to saturation. At saturation, increasing $\avg{n}$ (hence $\lambda_n$) increases the effective frequency, which pushes back the $d\tau_e/dJ=0$ threshold. Subsequently, revival of Rabi oscillations also help boost $\tau_e$.

\section{Quantum Circuitry for Flux Qubits}\label{sec.cir}
The discussion so far was general and did not pertain to any particular type of SC qubit. A specific design example is given in this section for the simplest flux qubits or rf-SQUIDs with $zz$ interactions. The qubits are capacitively coupled to a single resonator mode at the same time. Strong coupling of flux qubits to quantum oscillators has been experimentally demonstrated \cite{Chiorescu2004,Bourassa2009,Fedorov2010}. To exhibit quantum behavior the circuit's physical dimensions must be much smaller than the driving wavelength, which is easily achieved in these micron circuits\cite{Chiorescu2003}. Also tunable $zz$ interactions were experimentally realized for flux-qubits some years ago using inductive coupling\cite{Majer2005,vanderPloeg2007}. In general, various types of qubit-qubit interactions are possible for different qubit types\cite{Makhlin2001,Wendin2007,Geller2014}.

\begin{figure}
\includegraphics[width=0.75\columnwidth]{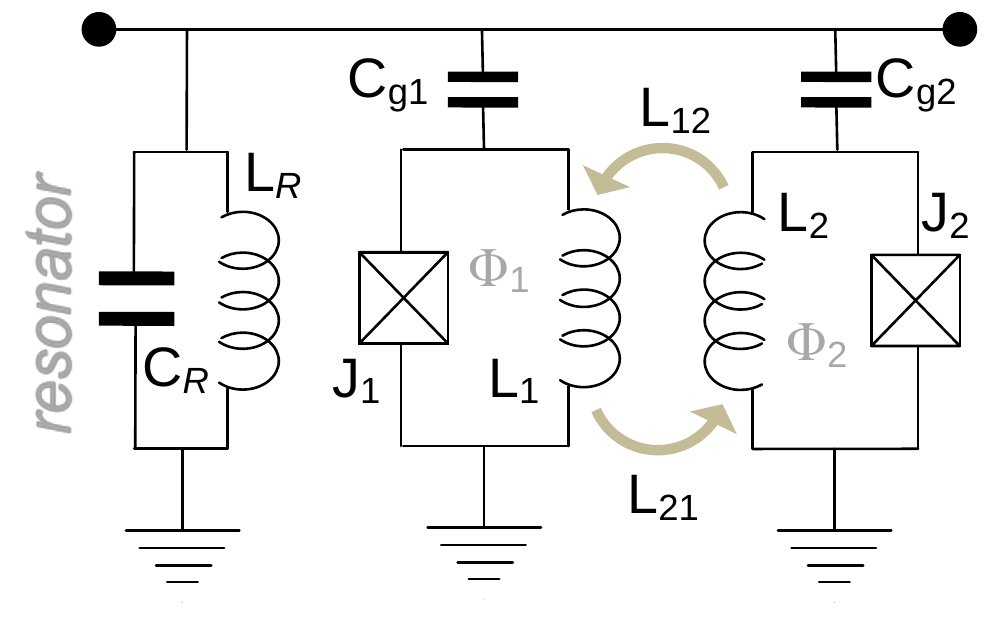}
\caption{(Color online) Circuit diagram for two SC flux qubits inductively coupled for $zz$ interactions, while also coupled to the single mode of a resonator or the LC oscillator.}
\label{fig:circuit}
\end{figure}

The Hamiltonian for the circuit in Fig.\ref{fig:circuit} consists of the following qubits, interaction and resonator terms
\begin{eqnarray}
\mathcal{H}=\mathcal{H_Q}+\mathcal{H_I}+\mathcal{H_R}.
\end{eqnarray}
In Fig.\ref{fig:circuit}, the potential energy term from the inductors is $\sum_{i,j}I_iL_{ij}I_j/2$ and the flux-linkage of loop-$j$ is $\Phi_j=\sum_iL_{ij}I_i$. From this and the potential energy of the JJ it follows that the Hamiltonian for the effective flux qubits with a single SC loop and a single JJ is:
\begin{equation}
\mathcal{H_Q}=\displaystyle\sum_{j=1,2}\frac{\hat Q_j^2}{2C'_j}-E^{(j)}_J\cos(\varphi^{(j)})-\frac{(\Phi^{(j)}-\Phi_e^{(j)})^2}{2L'_j}
\label{Hrf}
\end{equation}
where $\varphi^{(j)}=2\pi\Phi^{(j)}/\Phi_o$ is the phase difference across the JJ in loop-$j$, $\Phi_o=h/2e$ is the magnetic flux quantum, $\Phi^{(j)}$ and $\Phi^{(j)}_e$ are respectively the flux and the externally applied flux in the loop. Here $C'_j=C_j+C_{gj}$, $C_j$ is the junction capacitance and $L'_j=L_j-L_{ij}^2/L_i$ is the effective self inductance. It should be ensured that $L_{ij}\neq\sqrt{L_iL_j}$. The Josephson energy is $E_J=I_o\Phi_o/2\pi$ where, $I_o$ is the critical current. $\hat{Q}=-i\hbar\partial/\partial\Phi$ is the charge operator and $[\Phi,\hat{Q}]=i\hbar$.

\begin{figure}
\includegraphics[width=0.8\columnwidth]{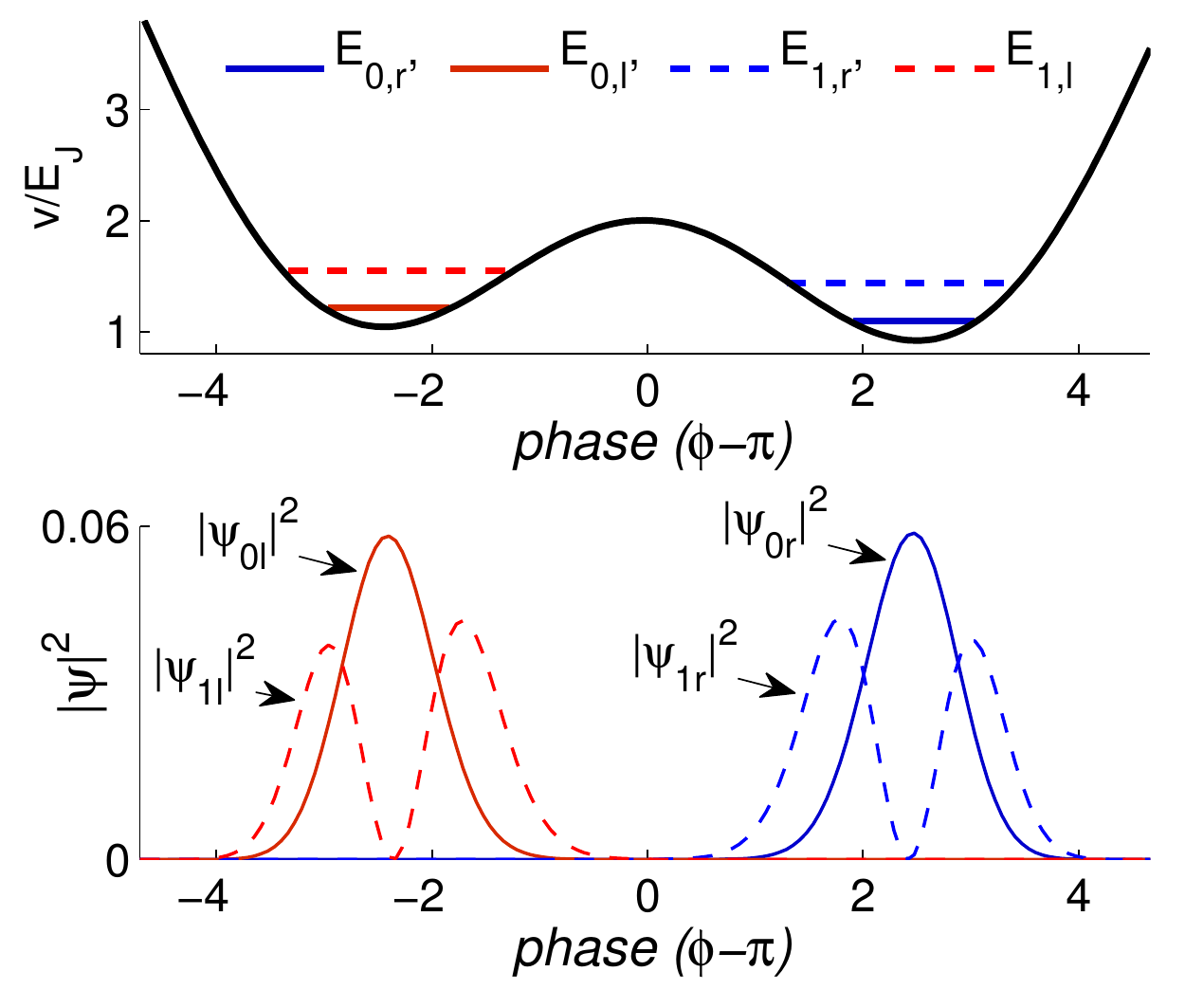}
\caption{(Color online) (top) Single qubit potential well and the ground state and the first excited state in each well (bottom) the respective wave functions. $\Psi_{0\ell}$ and $\Psi_{0 r}$ correspond to opposite directions of circulating persistent currents and form the flux qubit basis. Here the external $\varphi_{e}^{(j)} =\pi+\epsilon$, where $\epsilon$ breaks the degeneracy between the ground states $E_{0\ell}$ and $E_{0r}$.}
\label{fig:wav}
\end{figure}

In the long wavelength limit, using the canonical relations, the single qubit Hamiltonian can be mapped on to a tight-binding model:
\begin{equation}
\mathcal{H_{Q}}' = -\tau(c_{k}^{\dagger}c_{k+1} +  c_{k}^{\dagger}c_{k-1}) + (2\tau-{v}_{k})c_{k}^{\dagger}c_{k}
\label{eq:TB1Q}
\end{equation}
where $c_k$ and $c_k^{\dagger}$ are the $2e$ creation and annihilation operators for phase site-$k$, $\tau=E_C/a^2$, $E_C=(2e)^2/2C'_j$ is the JJ's charging energy, $a=\Delta\varphi$ is the lattice constant, $v_k= E_J\cos(\varphi_k)+ E_L(\varphi_k-\varphi_e)^2$ and $E_L=\hbar^2/2(2e)^2L'$.

For flux qubits $E_J>>E_C$. Sample calculations, using 100 gridsites and the lowest eigenenergies and eigenstates used to from the qubit basis is shown in fig.\ref{fig:wav}.
Here $\varphi_{e} =\pi+\epsilon$ (corresponding to the external flux: $\Phi_e\sim\Phi_o/2$) which gives the double well potential shown in the figure. $\epsilon$ breaks the degeneracy between the ground states $E_{0\ell}$ and $E_{0r}$. $\Psi_{0\ell}$ and $\Psi_{0 r}$ are the lowest states in each well, corresponding to opposite directions of circulating persistent currents and form the flux qubit basis as follows:
\begin{equation}
{\langle{\Psi_{0\ell},\Psi_{0r}}|\mathcal{H_Q}'|\Psi_{0\ell},\Psi_{0r}\rangle=\frac{\varepsilon}{2}\sigma_z+\frac{\delta}{2}\sigma_x},
\label{eq:PQ}
\end{equation}
where $\varepsilon=2I_p(\Phi_e-\Phi_o/2)$, $I_p$ is the persistent current in the loop and $\delta$ is the tunneling splitting that lifts the degeneracy between the clockwise and counterclockwise current states.

The qubit-qubit interaction term from considering the cross-flux-linkage in the loops in Fig.\ref{fig:circuit} is
\begin{equation}
\mathcal{H_I} = \frac{L_{12}}{L_{12}^2-L_1L_2}(\Phi_1-\Phi_{1e})(\Phi_2-\Phi_{2e})
\end{equation}
where $L_{12}=L_{21}$ is the mutual inductance, $\Phi_1=\Phi_kc_k^{\dagger}c_k\otimes I$ and $\Phi_2=I\otimes\Phi_kc_k^{\dagger}c_k$.
If the tunneling is negligible, $\delta\approx0$, then up to a minor phase factor
\begin{equation}
\bra{\Psi_{\mathcal Q}\otimes\Psi_{\mathcal Q}}\mathcal{H_I}\ket{\Psi_{\mathcal Q}\otimes\Psi_{\mathcal Q}} = J_{zz}\sigma_z\sigma_z
\end{equation}
where $\ket{\Psi_{\mathcal Q}}=|\Psi_{0\ell},\Psi_{0r}\rangle$. In the presence of tunneling there are small $\sigma_x\sigma_x$ terms and $\sigma_x\sigma_z$ type cross terms.

Note that in the presence of flux noise from the TLS in $\Phi_e$ and without tunneling, the projected interaction will also result in $J_{zz}[g^{(1)}_js^{(1)}_j(t)\times g^{(2)}_{j'}s^{(2)}_{j'}(t)]\sigma_z\sigma_z$ noise cross terms in Eq.\ref{Hn}, but these are very small.

Finally second quantizing the fundamental resonator mode, the resonator plus the qubit-resonator interaction term (coupled via $C_{gj}$) is
\begin{equation}
\mathcal{H_R} = \hbar\omega_r a^\dagger{a} + \sum_j \mathcal{V}_j\hat{Q}_j(a-a^\dagger)
\end{equation}
where $\omega_r=(L_RC_R)^{{-1/}{2}}$ and $\mathcal{V}_j=i(C_{gj}/C'_j)\sqrt{\hbar\omega_r/2C_R}$. Since $\hat{Q}=(c_{k}^{\dagger}c_{k+1} -  c_{k}^{\dagger}c_{k-1})\frac{e}{a}$, by projecting onto the qubit subspace for the lowest wavefunctions in fig.\ref{fig:wav} it is seen that $\bra{\Psi_{\mathcal Q}}\hat{Q}\ket{\Psi_{\mathcal Q}}\propto\sigma_y$. Therefore by making the RWA, $i\sigma_y(a-a^\dagger)\approx (\sigma_+a+\sigma_-a^{\dagger})$, the Jaynes-Cummings type coupling of Eq.\ref{Hjc2Q} is obtained.

\section{Comment on Lindblad Equations}\label{sec.Lind}
Lindblad master equations are one of the most commonly used methods for modeling dissipative open quantum systems. They dynamically map the density matrices in a completely positive and trace preserving way while assuming a memoryless quantum channel and have the general form:
\begin{equation}
\frac{d\rho}{dt} = -i[H,\rho] + \chi\rho{\chi^{\dagger}} -\frac{1}{2}\left( {\chi^{\dagger}}\chi\rho + \rho{\chi^{\dagger}}\chi   \right)
\label{Lind}
\end{equation}
Multiplying Eq.\ref{Lind} by the identity operator wherever necessary so that the identity \cite{ByronFuller.book}: $\overrightarrow{{\boldsymbol A}\rho {\boldsymbol B}} = ({\boldsymbol A}\otimes{\boldsymbol B})\vec{\rho}$ can be used to factor out $\vec{\rho}$, the Lindbald equation can be exactly transformed into a superoperator equation:
\begin{equation}
\frac{d\vec\rho}{dt} = \mathcal{L}\vec\rho
\label{SupOp}
\end{equation}
where $\vec{\rho}$ is a column vector representation of the density matrix,
$\mathcal{L}=\mathcal{L'}(H)+\mathcal{L^{\prime\prime}}(\chi)$ and
\begin{eqnarray}
\mathcal{L'}(H) &=& -iH\otimes{I} + iI\otimes{H}\\
\mathcal{L^{\prime\prime}}(\chi) &=& \chi\otimes{\chi^{\dagger}} -\frac{1}{2}\left( {\chi^{\dagger}}\chi\otimes{I} + {I}\otimes{\chi^{\dagger}}\chi\right)
\end{eqnarray}
Hence $\vec\rho(t) = \exp(\mathcal{L}t)\vec\rho(0)$.

Typically the qubit's relaxation and dephasing mechanisms are phenomenologically taken into account by introducing $T_1$ and $T_2$ times as follows:
\begin{equation}
\chi_q= \displaystyle\sum_j\frac{1}{\sqrt{T_{1j}}}\sigma^{(j)}_{-} + \sqrt{\frac{1}{2}\left(\frac{1}{T_{2j}} -  \frac{1}{2T_{1j}}\right)} \sigma^{(j)}_{z}.
\label{Xq}
\end{equation}
 While $\mathcal{L}'$ and $\mathcal{L^{\prime\prime}}$ do not necessarily commute, for the most general qubit-qubit interaction Hamiltonian ${H_I}=J_{\nu\nu}\sigma_\nu\sigma_\nu$, the corresponding superoperators commutes:
\begin{eqnarray}
\big[ \mathcal{L'}({H_I}), \mathcal{L^{\prime\prime}}(\chi_q)\big]=0.
\label{Comm}
\end{eqnarray}
Hence the interaction term will have absolutely no effect on the dynamics and therefore the interaction-driven-suppression-of-decoherence is not reproducible with the Lindblad formalism. This is because the Lindblad method assumes delta function correlations for the noise\cite{Blum.book}. Whereas an accurate treatment of $1/f$ noise requires $e^{-2\gamma t}$ type correlations with a log-normal distribution of $\gamma$. The quasi Hamiltonian method treats this noise correlation distribution exactly. Also, the Lindblad equations are derived by expanding the von-Neumann equation(VNE) to second order, whereas the quasi-Hamimtonian method solves the VNE and the stochastic master equation exactly.

The decay of the fundamental resonator mode is similarly modeled by considering $\chi_r= \sqrt{\omega_r/{\rm{Q}}}a$,
where ${\rm Q}$ is the resonator's quality factor.
Within the Lindblad formalism the interaction term does not affect this decay for the reasons mentioned above.
Overall a high quality factor resonator would be best for the suggested effects to be observable. Alternatively the detuning, $\Delta$, could also be a useful parameter.
Very high $\rm Q$-factors have been reported for coplanar waveguide resonators hosting flux qubits\cite{Niemczyk2010,Xiang2013RMP}. Newer, high-${\rm Q}$ 3D resonator cavities, such as those used for transmons\cite{Paik2011} and flux qubits\cite{Stern2014}, could also be very useful.


\section{Summary}\label{sec.sum}
As long as the qubit-qubit interactions are orthogonal to the resonator-qubit coupling, the coherence times are shown to improve drastically for the entangled Bell states, where each physical qubit subjected to local low frequency noise. These improvements however tend to saturate and then decrease slowly with further increases in the interaction strength. At this saturation point the resonator amplifies the normalized coherence enhancements when loaded with more photons. When compared to $\langle{n}\rangle=1$, the resonator can always amplify the absolute enhancement. Coupling between the interactions and the resonator mode also induce unusual collapse and revival of the entanglement dynamics.

While the $zz$-interactions are optimal for boosting the coherence times, the proof-of-concept could also be established with $yy$ interactions. Overall, the quasi-Hamiltonian treatment of the Tavis-Cummings model with the TLS is exact and is key in revealing these results -- which cannot be obtained with other methods that make the Markov approximation.

\section{Acknowledgements}

I wish to thank Robert Joynt for introducing me to the quasi Hamiltonian method and for the many invaluable interactions. In addition I would like to thank Robert McDermott, Leonid Pryadko, Alexander Korotkov, and Oney Soykal for several very helpful discussions. AD is supported by the U.S. Army Research Office, Grant No.~W911NF-11-1-0027, and by the NSF, Grant No.~1018935,
and previously by DARPA-QuEst Grant No. MSN118850.

\bibliographystyle{apsrev}

\begin{thebibliography}{78}
\expandafter\ifx\csname natexlab\endcsname\relax\def\natexlab#1{#1}\fi
\expandafter\ifx\csname bibnamefont\endcsname\relax
  \def\bibnamefont#1{#1}\fi
\expandafter\ifx\csname bibfnamefont\endcsname\relax
  \def\bibfnamefont#1{#1}\fi
\expandafter\ifx\csname citenamefont\endcsname\relax
  \def\citenamefont#1{#1}\fi
\expandafter\ifx\csname url\endcsname\relax
  \def\url#1{\texttt{#1}}\fi
\expandafter\ifx\csname urlprefix\endcsname\relax\def\urlprefix{URL }\fi
\providecommand{\bibinfo}[2]{#2}
\providecommand{\eprint}[2][]{\url{#2}}

\bibitem[{\citenamefont{Raimond et~al.}(2001)\citenamefont{Raimond, Brune, and
  Haroche}}]{Raimond2001}
\bibinfo{author}{\bibfnamefont{J.~M.} \bibnamefont{Raimond}},
  \bibinfo{author}{\bibfnamefont{M.}~\bibnamefont{Brune}}, \bibnamefont{and}
  \bibinfo{author}{\bibfnamefont{S.}~\bibnamefont{Haroche}},
  \bibinfo{journal}{Rev. Mod. Phys.} \textbf{\bibinfo{volume}{73}},
  \bibinfo{pages}{565} (\bibinfo{year}{2001}).

\bibitem[{\citenamefont{Leibfried et~al.}(2003)\citenamefont{Leibfried, Blatt,
  Monroe, and Wineland}}]{Leibfried2003}
\bibinfo{author}{\bibfnamefont{D.}~\bibnamefont{Leibfried}},
  \bibinfo{author}{\bibfnamefont{R.}~\bibnamefont{Blatt}},
  \bibinfo{author}{\bibfnamefont{C.}~\bibnamefont{Monroe}}, \bibnamefont{and}
  \bibinfo{author}{\bibfnamefont{D.}~\bibnamefont{Wineland}},
  \bibinfo{journal}{Rev. Mod. Phys.} \textbf{\bibinfo{volume}{75}},
  \bibinfo{pages}{281} (\bibinfo{year}{2003}).

\bibitem[{\citenamefont{Hennessy et~al.}(2007)\citenamefont{Hennessy, Badolato,
  Winger, Gerace, Atature, Gulde, Falt, Hu, and Imamoglu}}]{Hennessy2007}
\bibinfo{author}{\bibfnamefont{K.}~\bibnamefont{Hennessy}},
  \bibinfo{author}{\bibfnamefont{A.}~\bibnamefont{Badolato}},
  \bibinfo{author}{\bibfnamefont{M.}~\bibnamefont{Winger}},
  \bibinfo{author}{\bibfnamefont{D.}~\bibnamefont{Gerace}},
  \bibinfo{author}{\bibfnamefont{M.}~\bibnamefont{Atature}},
  \bibinfo{author}{\bibfnamefont{S.}~\bibnamefont{Gulde}},
  \bibinfo{author}{\bibfnamefont{S.}~\bibnamefont{Falt}},
  \bibinfo{author}{\bibfnamefont{E.~L.} \bibnamefont{Hu}}, \bibnamefont{and}
  \bibinfo{author}{\bibfnamefont{A.}~\bibnamefont{Imamoglu}},
  \bibinfo{journal}{Nature} \textbf{\bibinfo{volume}{445}},
  \bibinfo{pages}{896} (\bibinfo{year}{2007}).

\bibitem[{\citenamefont{Devoret et~al.}(1989)\citenamefont{Devoret, Esteve,
  Martinis, and Urbina}}]{Devoret1989}
\bibinfo{author}{\bibfnamefont{M.~H.} \bibnamefont{Devoret}},
  \bibinfo{author}{\bibfnamefont{D.}~\bibnamefont{Esteve}},
  \bibinfo{author}{\bibfnamefont{J.~M.} \bibnamefont{Martinis}},
  \bibnamefont{and} \bibinfo{author}{\bibfnamefont{C.}~\bibnamefont{Urbina}},
  \bibinfo{journal}{Physica Scripta} \textbf{\bibinfo{volume}{1989}},
  \bibinfo{pages}{118} (\bibinfo{year}{1989}).

\bibitem[{\citenamefont{Schoelkopf and Girvin}(2008)}]{Schoelkopf2008}
\bibinfo{author}{\bibfnamefont{R.~J.} \bibnamefont{Schoelkopf}}
  \bibnamefont{and} \bibinfo{author}{\bibfnamefont{S.~M.}
  \bibnamefont{Girvin}}, \bibinfo{journal}{Nature}
  \textbf{\bibinfo{volume}{451}}, \bibinfo{pages}{664} (\bibinfo{year}{2008}).

\bibitem[{\citenamefont{Martinis}(2009)}]{Martinis2009}
\bibinfo{author}{\bibfnamefont{J.~M.} \bibnamefont{Martinis}},
  \bibinfo{journal}{Quantum Information Processing}
  \textbf{\bibinfo{volume}{8}}, \bibinfo{pages}{81} (\bibinfo{year}{2009}),
  ISSN \bibinfo{issn}{1570-0755}.

\bibitem[{\citenamefont{Niemczyk et~al.}(2010)\citenamefont{Niemczyk, Deppe,
  Huebl, Menzel, Hocke, Schwarz, Garcia-Ripoll, Zueco, Hummer, Solano
  et~al.}}]{Niemczyk2010}
\bibinfo{author}{\bibfnamefont{T.}~\bibnamefont{Niemczyk}},
  \bibinfo{author}{\bibfnamefont{F.}~\bibnamefont{Deppe}},
  \bibinfo{author}{\bibfnamefont{H.}~\bibnamefont{Huebl}},
  \bibinfo{author}{\bibfnamefont{E.~P.} \bibnamefont{Menzel}},
  \bibinfo{author}{\bibfnamefont{F.}~\bibnamefont{Hocke}},
  \bibinfo{author}{\bibfnamefont{M.~J.} \bibnamefont{Schwarz}},
  \bibinfo{author}{\bibfnamefont{J.~J.} \bibnamefont{Garcia-Ripoll}},
  \bibinfo{author}{\bibfnamefont{D.}~\bibnamefont{Zueco}},
  \bibinfo{author}{\bibfnamefont{T.}~\bibnamefont{Hummer}},
  \bibinfo{author}{\bibfnamefont{E.}~\bibnamefont{Solano}},
  \bibnamefont{et~al.}, \bibinfo{journal}{Nat. Phys.}
  \textbf{\bibinfo{volume}{6}}, \bibinfo{pages}{772} (\bibinfo{year}{2010}).

\bibitem[{\citenamefont{Blais et~al.}(2004)\citenamefont{Blais, Huang,
  Wallraff, Girvin, and Schoelkopf}}]{Blais2004}
\bibinfo{author}{\bibfnamefont{A.}~\bibnamefont{Blais}},
  \bibinfo{author}{\bibfnamefont{R.-S.} \bibnamefont{Huang}},
  \bibinfo{author}{\bibfnamefont{A.}~\bibnamefont{Wallraff}},
  \bibinfo{author}{\bibfnamefont{S.~M.} \bibnamefont{Girvin}},
  \bibnamefont{and} \bibinfo{author}{\bibfnamefont{R.~J.}
  \bibnamefont{Schoelkopf}}, \bibinfo{journal}{Phys. Rev. A}
  \textbf{\bibinfo{volume}{69}}, \bibinfo{pages}{062320}
  (\bibinfo{year}{2004}).

\bibitem[{\citenamefont{Clarke and Wilhelm}(2008)}]{Clarke2008}
\bibinfo{author}{\bibfnamefont{J.}~\bibnamefont{Clarke}} \bibnamefont{and}
  \bibinfo{author}{\bibfnamefont{F.~K.} \bibnamefont{Wilhelm}},
  \bibinfo{journal}{Nature} \textbf{\bibinfo{volume}{453}},
  \bibinfo{pages}{1031} (\bibinfo{year}{2008}).

\bibitem[{\citenamefont{Yoshihara et~al.}(2006)\citenamefont{Yoshihara,
  Harrabi, Niskanen, Nakamura, and Tsai}}]{Yoshihara2006}
\bibinfo{author}{\bibfnamefont{F.}~\bibnamefont{Yoshihara}},
  \bibinfo{author}{\bibfnamefont{K.}~\bibnamefont{Harrabi}},
  \bibinfo{author}{\bibfnamefont{A.~O.} \bibnamefont{Niskanen}},
  \bibinfo{author}{\bibfnamefont{Y.}~\bibnamefont{Nakamura}}, \bibnamefont{and}
  \bibinfo{author}{\bibfnamefont{J.~S.} \bibnamefont{Tsai}},
  \bibinfo{journal}{Phys. Rev. Lett.} \textbf{\bibinfo{volume}{97}},
  \bibinfo{pages}{167001} (\bibinfo{year}{2006}).

\bibitem[{\citenamefont{Sank et~al.}(2012)\citenamefont{Sank, Barends,
  Bialczak, Chen, Kelly, Lenander, Lucero, Mariantoni, Megrant, Neeley
  et~al.}}]{Sank2012}
\bibinfo{author}{\bibfnamefont{D.}~\bibnamefont{Sank}},
  \bibinfo{author}{\bibfnamefont{R.}~\bibnamefont{Barends}},
  \bibinfo{author}{\bibfnamefont{R.~C.} \bibnamefont{Bialczak}},
  \bibinfo{author}{\bibfnamefont{Y.}~\bibnamefont{Chen}},
  \bibinfo{author}{\bibfnamefont{J.}~\bibnamefont{Kelly}},
  \bibinfo{author}{\bibfnamefont{M.}~\bibnamefont{Lenander}},
  \bibinfo{author}{\bibfnamefont{E.}~\bibnamefont{Lucero}},
  \bibinfo{author}{\bibfnamefont{M.}~\bibnamefont{Mariantoni}},
  \bibinfo{author}{\bibfnamefont{A.}~\bibnamefont{Megrant}},
  \bibinfo{author}{\bibfnamefont{M.}~\bibnamefont{Neeley}},
  \bibnamefont{et~al.}, \bibinfo{journal}{Phys. Rev. Lett.}
  \textbf{\bibinfo{volume}{109}}, \bibinfo{pages}{067001}
  (\bibinfo{year}{2012}).

\bibitem[{\citenamefont{Ithier et~al.}(2005)\citenamefont{Ithier, Collin,
  Joyez, Meeson, Vion, Esteve, Chiarello, Shnirman, Makhlin, Schriefl
  et~al.}}]{Ithier2005}
\bibinfo{author}{\bibfnamefont{G.}~\bibnamefont{Ithier}},
  \bibinfo{author}{\bibfnamefont{E.}~\bibnamefont{Collin}},
  \bibinfo{author}{\bibfnamefont{P.}~\bibnamefont{Joyez}},
  \bibinfo{author}{\bibfnamefont{P.~J.} \bibnamefont{Meeson}},
  \bibinfo{author}{\bibfnamefont{D.}~\bibnamefont{Vion}},
  \bibinfo{author}{\bibfnamefont{D.}~\bibnamefont{Esteve}},
  \bibinfo{author}{\bibfnamefont{F.}~\bibnamefont{Chiarello}},
  \bibinfo{author}{\bibfnamefont{A.}~\bibnamefont{Shnirman}},
  \bibinfo{author}{\bibfnamefont{Y.}~\bibnamefont{Makhlin}},
  \bibinfo{author}{\bibfnamefont{J.}~\bibnamefont{Schriefl}},
  \bibnamefont{et~al.}, \bibinfo{journal}{Phys. Rev. B}
  \textbf{\bibinfo{volume}{72}}, \bibinfo{pages}{134519}
  (\bibinfo{year}{2005}).

\bibitem[{\citenamefont{McDermott}(2009)}]{McDermott2009}
\bibinfo{author}{\bibfnamefont{R.}~\bibnamefont{McDermott}},
  \bibinfo{journal}{Applied Superconductivity, IEEE Transactions on}
  \textbf{\bibinfo{volume}{19}}, \bibinfo{pages}{2 } (\bibinfo{year}{2009}),
  ISSN \bibinfo{issn}{1051-8223}.

\bibitem[{\citenamefont{Paladino et~al.}(2014)\citenamefont{Paladino, Galperin,
  Falci, and Altshuler}}]{Paladino2014}
\bibinfo{author}{\bibfnamefont{E.}~\bibnamefont{Paladino}},
  \bibinfo{author}{\bibfnamefont{M.}~\bibnamefont{Galperin},
  \bibfnamefont{Y.}}, \bibinfo{author}{\bibfnamefont{G.}~\bibnamefont{Falci}},
  \bibnamefont{and} \bibinfo{author}{\bibfnamefont{L.}~\bibnamefont{Altshuler},
  \bibfnamefont{B.}}, \bibinfo{journal}{Rev. Mod. Phys.}
  \textbf{\bibinfo{volume}{86}}, \bibinfo{pages}{361} (\bibinfo{year}{2014}).

\bibitem[{\citenamefont{Koch et~al.}(2007{\natexlab{a}})\citenamefont{Koch, Yu,
  Gambetta, Houck, Schuster, Majer, Blais, Devoret, Girvin, and
  Schoelkopf}}]{Koch2007pra}
\bibinfo{author}{\bibfnamefont{J.}~\bibnamefont{Koch}},
  \bibinfo{author}{\bibfnamefont{T.~M.} \bibnamefont{Yu}},
  \bibinfo{author}{\bibfnamefont{J.}~\bibnamefont{Gambetta}},
  \bibinfo{author}{\bibfnamefont{A.~A.} \bibnamefont{Houck}},
  \bibinfo{author}{\bibfnamefont{D.~I.} \bibnamefont{Schuster}},
  \bibinfo{author}{\bibfnamefont{J.}~\bibnamefont{Majer}},
  \bibinfo{author}{\bibfnamefont{A.}~\bibnamefont{Blais}},
  \bibinfo{author}{\bibfnamefont{M.~H.} \bibnamefont{Devoret}},
  \bibinfo{author}{\bibfnamefont{S.~M.} \bibnamefont{Girvin}},
  \bibnamefont{and} \bibinfo{author}{\bibfnamefont{R.~J.}
  \bibnamefont{Schoelkopf}}, \bibinfo{journal}{Phys. Rev. A}
  \textbf{\bibinfo{volume}{76}}, \bibinfo{pages}{042319}
  (\bibinfo{year}{2007}{\natexlab{a}}).

\bibitem[{\citenamefont{Koch et~al.}(1983)\citenamefont{Koch, Clarke, Goubau,
  Martinis, Pegrum, and Harlingen}}]{Koch1983}
\bibinfo{author}{\bibfnamefont{R.~H.} \bibnamefont{Koch}},
  \bibinfo{author}{\bibfnamefont{J.}~\bibnamefont{Clarke}},
  \bibinfo{author}{\bibfnamefont{W.~M.} \bibnamefont{Goubau}},
  \bibinfo{author}{\bibfnamefont{J.~M.} \bibnamefont{Martinis}},
  \bibinfo{author}{\bibfnamefont{C.~M.} \bibnamefont{Pegrum}},
  \bibnamefont{and} \bibinfo{author}{\bibfnamefont{D.~J.}
  \bibnamefont{Harlingen}}, \bibinfo{journal}{Journal of Low Temperature
  Physics} \textbf{\bibinfo{volume}{51}}, \bibinfo{pages}{207}
  (\bibinfo{year}{1983}), ISSN \bibinfo{issn}{0022-2291},
  \bibinfo{note}{10.1007/BF00683423}.

\bibitem[{\citenamefont{Wellstood et~al.}(1987)\citenamefont{Wellstood, Urbina,
  and Clarke}}]{Wellstood1987}
\bibinfo{author}{\bibfnamefont{F.}~\bibnamefont{Wellstood}},
  \bibinfo{author}{\bibfnamefont{C.}~\bibnamefont{Urbina}}, \bibnamefont{and}
  \bibinfo{author}{\bibfnamefont{J.}~\bibnamefont{Clarke}},
  \bibinfo{journal}{Magnetics, IEEE Transactions on}
  \textbf{\bibinfo{volume}{23}}, \bibinfo{pages}{1662 } (\bibinfo{year}{1987}),
  ISSN \bibinfo{issn}{0018-9464}.

\bibitem[{\citenamefont{Bialczak et~al.}(2007)\citenamefont{Bialczak,
  McDermott, Ansmann, Hofheinz, Katz, Lucero, Neeley, O'Connell, Wang, Cleland
  et~al.}}]{Bialczak2007}
\bibinfo{author}{\bibfnamefont{R.~C.} \bibnamefont{Bialczak}},
  \bibinfo{author}{\bibfnamefont{R.}~\bibnamefont{McDermott}},
  \bibinfo{author}{\bibfnamefont{M.}~\bibnamefont{Ansmann}},
  \bibinfo{author}{\bibfnamefont{M.}~\bibnamefont{Hofheinz}},
  \bibinfo{author}{\bibfnamefont{N.}~\bibnamefont{Katz}},
  \bibinfo{author}{\bibfnamefont{E.}~\bibnamefont{Lucero}},
  \bibinfo{author}{\bibfnamefont{M.}~\bibnamefont{Neeley}},
  \bibinfo{author}{\bibfnamefont{A.~D.} \bibnamefont{O'Connell}},
  \bibinfo{author}{\bibfnamefont{H.}~\bibnamefont{Wang}},
  \bibinfo{author}{\bibfnamefont{A.~N.} \bibnamefont{Cleland}},
  \bibnamefont{et~al.}, \bibinfo{journal}{Phys. Rev. Lett.}
  \textbf{\bibinfo{volume}{99}}, \bibinfo{pages}{187006}
  (\bibinfo{year}{2007}).

\bibitem[{\citenamefont{Koch et~al.}(2007{\natexlab{b}})\citenamefont{Koch,
  DiVincenzo, and Clarke}}]{Koch2007}
\bibinfo{author}{\bibfnamefont{R.~H.} \bibnamefont{Koch}},
  \bibinfo{author}{\bibfnamefont{D.~P.} \bibnamefont{DiVincenzo}},
  \bibnamefont{and} \bibinfo{author}{\bibfnamefont{J.}~\bibnamefont{Clarke}},
  \bibinfo{journal}{Phys. Rev. Lett.} \textbf{\bibinfo{volume}{98}},
  \bibinfo{pages}{267003} (\bibinfo{year}{2007}{\natexlab{b}}).

\bibitem[{\citenamefont{de~Sousa}(2007)}]{deSousa2007}
\bibinfo{author}{\bibfnamefont{R.}~\bibnamefont{de~Sousa}},
  \bibinfo{journal}{Phys. Rev. B} \textbf{\bibinfo{volume}{76}},
  \bibinfo{pages}{245306} (\bibinfo{year}{2007}).

\bibitem[{\citenamefont{Faoro and Ioffe}(2008)}]{Faoro2008}
\bibinfo{author}{\bibfnamefont{L.}~\bibnamefont{Faoro}} \bibnamefont{and}
  \bibinfo{author}{\bibfnamefont{L.~B.} \bibnamefont{Ioffe}},
  \bibinfo{journal}{Phys. Rev. Lett.} \textbf{\bibinfo{volume}{100}},
  \bibinfo{pages}{227005} (\bibinfo{year}{2008}).

\bibitem[{\citenamefont{Choi et~al.}(2009)\citenamefont{Choi, Lee, Louie, and
  Clarke}}]{Choi2009}
\bibinfo{author}{\bibfnamefont{S.}~\bibnamefont{Choi}},
  \bibinfo{author}{\bibfnamefont{D.-H.} \bibnamefont{Lee}},
  \bibinfo{author}{\bibfnamefont{S.~G.} \bibnamefont{Louie}}, \bibnamefont{and}
  \bibinfo{author}{\bibfnamefont{J.}~\bibnamefont{Clarke}},
  \bibinfo{journal}{Phys. Rev. Lett.} \textbf{\bibinfo{volume}{103}},
  \bibinfo{pages}{197001} (\bibinfo{year}{2009}).

\bibitem[{\citenamefont{Chen and Yu}(2010)}]{Chen2010}
\bibinfo{author}{\bibfnamefont{Z.}~\bibnamefont{Chen}} \bibnamefont{and}
  \bibinfo{author}{\bibfnamefont{C.~C.} \bibnamefont{Yu}},
  \bibinfo{journal}{Phys. Rev. Lett.} \textbf{\bibinfo{volume}{104}},
  \bibinfo{pages}{247204} (\bibinfo{year}{2010}).

\bibitem[{\citenamefont{K.~Kechedzhi and Ioffe}(2011)}]{Kechedzhi2011}
\bibinfo{author}{\bibfnamefont{L.~F.} \bibnamefont{K.~Kechedzhi}}
  \bibnamefont{and} \bibinfo{author}{\bibfnamefont{L.~B.} \bibnamefont{Ioffe}},
  \bibinfo{journal}{ArXiv} p. \bibinfo{pages}{1102.3445}
  (\bibinfo{year}{2011}).

\bibitem[{\citenamefont{{De}}(2014)}]{De2014arxiv}
\bibinfo{author}{\bibfnamefont{A.}~\bibnamefont{{De}}}, \bibinfo{journal}{ArXiv
  e-prints}  (\bibinfo{year}{2014}), \eprint{1403.0124}.

\bibitem[{\citenamefont{Atalaya et~al.}(2014)\citenamefont{Atalaya, Clarke,
  Sch{\"o}n, and Shnirman}}]{Atalaya2014}
\bibinfo{author}{\bibfnamefont{J.}~\bibnamefont{Atalaya}},
  \bibinfo{author}{\bibfnamefont{J.}~\bibnamefont{Clarke}},
  \bibinfo{author}{\bibfnamefont{G.}~\bibnamefont{Sch{\"o}n}},
  \bibnamefont{and} \bibinfo{author}{\bibfnamefont{A.}~\bibnamefont{Shnirman}},
  \bibinfo{journal}{arXiv:1402.6229}  (\bibinfo{year}{2014}).

\bibitem[{\citenamefont{Sendelbach et~al.}(2008)\citenamefont{Sendelbach,
  Hover, Kittel, M\"uck, Martinis, and McDermott}}]{Sendelbach2008}
\bibinfo{author}{\bibfnamefont{S.}~\bibnamefont{Sendelbach}},
  \bibinfo{author}{\bibfnamefont{D.}~\bibnamefont{Hover}},
  \bibinfo{author}{\bibfnamefont{A.}~\bibnamefont{Kittel}},
  \bibinfo{author}{\bibfnamefont{M.}~\bibnamefont{M\"uck}},
  \bibinfo{author}{\bibfnamefont{J.~M.} \bibnamefont{Martinis}},
  \bibnamefont{and}
  \bibinfo{author}{\bibfnamefont{R.}~\bibnamefont{McDermott}},
  \bibinfo{journal}{Phys. Rev. Lett.} \textbf{\bibinfo{volume}{100}},
  \bibinfo{pages}{227006} (\bibinfo{year}{2008}).

\bibitem[{\citenamefont{Bluhm et~al.}(2009)\citenamefont{Bluhm, Bert, Koshnick,
  Huber, and Moler}}]{Bluhm2009}
\bibinfo{author}{\bibfnamefont{H.}~\bibnamefont{Bluhm}},
  \bibinfo{author}{\bibfnamefont{J.~A.} \bibnamefont{Bert}},
  \bibinfo{author}{\bibfnamefont{N.~C.} \bibnamefont{Koshnick}},
  \bibinfo{author}{\bibfnamefont{M.~E.} \bibnamefont{Huber}}, \bibnamefont{and}
  \bibinfo{author}{\bibfnamefont{K.~A.} \bibnamefont{Moler}},
  \bibinfo{journal}{Phys. Rev. Lett.} \textbf{\bibinfo{volume}{103}},
  \bibinfo{pages}{026805} (\bibinfo{year}{2009}).

\bibitem[{\citenamefont{Anton et~al.}(2013)\citenamefont{Anton, Birenbaum,
  O'Kelley, Bolkhovsky, Braje, Fitch, Neeley, Hilton, Cho, Irwin
  et~al.}}]{Anton2013}
\bibinfo{author}{\bibfnamefont{S.~M.} \bibnamefont{Anton}},
  \bibinfo{author}{\bibfnamefont{J.~S.} \bibnamefont{Birenbaum}},
  \bibinfo{author}{\bibfnamefont{S.~R.} \bibnamefont{O'Kelley}},
  \bibinfo{author}{\bibfnamefont{V.}~\bibnamefont{Bolkhovsky}},
  \bibinfo{author}{\bibfnamefont{D.~A.} \bibnamefont{Braje}},
  \bibinfo{author}{\bibfnamefont{G.}~\bibnamefont{Fitch}},
  \bibinfo{author}{\bibfnamefont{M.}~\bibnamefont{Neeley}},
  \bibinfo{author}{\bibfnamefont{G.~C.} \bibnamefont{Hilton}},
  \bibinfo{author}{\bibfnamefont{H.-M.} \bibnamefont{Cho}},
  \bibinfo{author}{\bibfnamefont{K.~D.} \bibnamefont{Irwin}},
  \bibnamefont{et~al.}, \bibinfo{journal}{Phys. Rev. Lett.}
  \textbf{\bibinfo{volume}{110}}, \bibinfo{pages}{147002}
  (\bibinfo{year}{2013}).

\bibitem[{\citenamefont{Martinis et~al.}(2005)\citenamefont{Martinis, Cooper,
  McDermott, Steffen, Ansmann, Osborn, Cicak, Oh, Pappas, Simmonds
  et~al.}}]{Martinis2005}
\bibinfo{author}{\bibfnamefont{J.~M.} \bibnamefont{Martinis}},
  \bibinfo{author}{\bibfnamefont{K.~B.} \bibnamefont{Cooper}},
  \bibinfo{author}{\bibfnamefont{R.}~\bibnamefont{McDermott}},
  \bibinfo{author}{\bibfnamefont{M.}~\bibnamefont{Steffen}},
  \bibinfo{author}{\bibfnamefont{M.}~\bibnamefont{Ansmann}},
  \bibinfo{author}{\bibfnamefont{K.~D.} \bibnamefont{Osborn}},
  \bibinfo{author}{\bibfnamefont{K.}~\bibnamefont{Cicak}},
  \bibinfo{author}{\bibfnamefont{S.}~\bibnamefont{Oh}},
  \bibinfo{author}{\bibfnamefont{D.~P.} \bibnamefont{Pappas}},
  \bibinfo{author}{\bibfnamefont{R.~W.} \bibnamefont{Simmonds}},
  \bibnamefont{et~al.}, \bibinfo{journal}{Phys. Rev. Lett.}
  \textbf{\bibinfo{volume}{95}}, \bibinfo{pages}{210503}
  (\bibinfo{year}{2005}).

\bibitem[{\citenamefont{Oh et~al.}(2006)\citenamefont{Oh, Cicak, Kline,
  Sillanp\"a\"a, Osborn, Whittaker, Simmonds, and Pappas}}]{Oh2006}
\bibinfo{author}{\bibfnamefont{S.}~\bibnamefont{Oh}},
  \bibinfo{author}{\bibfnamefont{K.}~\bibnamefont{Cicak}},
  \bibinfo{author}{\bibfnamefont{J.~S.} \bibnamefont{Kline}},
  \bibinfo{author}{\bibfnamefont{M.~A.} \bibnamefont{Sillanp\"a\"a}},
  \bibinfo{author}{\bibfnamefont{K.~D.} \bibnamefont{Osborn}},
  \bibinfo{author}{\bibfnamefont{J.~D.} \bibnamefont{Whittaker}},
  \bibinfo{author}{\bibfnamefont{R.~W.} \bibnamefont{Simmonds}},
  \bibnamefont{and} \bibinfo{author}{\bibfnamefont{D.~P.}
  \bibnamefont{Pappas}}, \bibinfo{journal}{Phys. Rev. B}
  \textbf{\bibinfo{volume}{74}}, \bibinfo{pages}{100502}
  (\bibinfo{year}{2006}).

\bibitem[{\citenamefont{Khodjasteh and Lidar}(2005)}]{Khodjasteh2005}
\bibinfo{author}{\bibfnamefont{K.}~\bibnamefont{Khodjasteh}} \bibnamefont{and}
  \bibinfo{author}{\bibfnamefont{D.~A.} \bibnamefont{Lidar}},
  \bibinfo{journal}{Phys. Rev. Lett.} \textbf{\bibinfo{volume}{95}},
  \bibinfo{pages}{180501} (\bibinfo{year}{2005}).

\bibitem[{\citenamefont{Uhrig}(2007)}]{Uhrig2007}
\bibinfo{author}{\bibfnamefont{G.~S.} \bibnamefont{Uhrig}},
  \bibinfo{journal}{Phys. Rev. Lett.} \textbf{\bibinfo{volume}{98}},
  \bibinfo{pages}{100504} (\bibinfo{year}{2007}).

\bibitem[{\citenamefont{Kitaev}(2001)}]{Kitaev:2001}
\bibinfo{author}{\bibfnamefont{A.~Y.} \bibnamefont{Kitaev}},
  \bibinfo{journal}{Physics-Uspekhi} \textbf{\bibinfo{volume}{44}},
  \bibinfo{pages}{131} (\bibinfo{year}{2001}).

\bibitem[{\citenamefont{Bacon et~al.}(2000)\citenamefont{Bacon, Kempe, Lidar,
  and Whaley}}]{Bacon2000}
\bibinfo{author}{\bibfnamefont{D.}~\bibnamefont{Bacon}},
  \bibinfo{author}{\bibfnamefont{J.}~\bibnamefont{Kempe}},
  \bibinfo{author}{\bibfnamefont{D.~A.} \bibnamefont{Lidar}}, \bibnamefont{and}
  \bibinfo{author}{\bibfnamefont{K.~B.} \bibnamefont{Whaley}},
  \bibinfo{journal}{Phys. Rev. Lett.} \textbf{\bibinfo{volume}{85}},
  \bibinfo{pages}{1758} (\bibinfo{year}{2000}).

\bibitem[{\citenamefont{Lidar}(2008)}]{Lidar2008}
\bibinfo{author}{\bibfnamefont{D.~A.} \bibnamefont{Lidar}},
  \bibinfo{journal}{Phys. Rev. Lett.} \textbf{\bibinfo{volume}{100}},
  \bibinfo{pages}{160506} (\bibinfo{year}{2008}).

\bibitem[{\citenamefont{Calderbank and Shor}(1996)}]{Calderbank1996}
\bibinfo{author}{\bibfnamefont{A.~R.} \bibnamefont{Calderbank}}
  \bibnamefont{and} \bibinfo{author}{\bibfnamefont{P.~W.} \bibnamefont{Shor}},
  \bibinfo{journal}{Phys. Rev. A} \textbf{\bibinfo{volume}{54}},
  \bibinfo{pages}{1098} (\bibinfo{year}{1996}).

\bibitem[{\citenamefont{Steane}(1996)}]{Steane1996}
\bibinfo{author}{\bibfnamefont{A.~M.} \bibnamefont{Steane}},
  \bibinfo{journal}{Phys. Rev. Lett.} \textbf{\bibinfo{volume}{77}},
  \bibinfo{pages}{793} (\bibinfo{year}{1996}).

\bibitem[{\citenamefont{Knill et~al.}(1998)\citenamefont{Knill, Laflamme, and
  Zurek}}]{Knill-error-bound2}
\bibinfo{author}{\bibfnamefont{E.}~\bibnamefont{Knill}},
  \bibinfo{author}{\bibfnamefont{R.}~\bibnamefont{Laflamme}}, \bibnamefont{and}
  \bibinfo{author}{\bibfnamefont{W.~H.} \bibnamefont{Zurek}},
  \bibinfo{journal}{Science} \textbf{\bibinfo{volume}{279}},
  \bibinfo{pages}{342} (\bibinfo{year}{1998}).

\bibitem[{\citenamefont{Vandersypen et~al.}(2004)\citenamefont{Vandersypen,
  Elzerman, Schouten, van Beveren, Hanson, and Kouwenhoven}}]{Vandersypen2004}
\bibinfo{author}{\bibfnamefont{L.~M.~K.} \bibnamefont{Vandersypen}},
  \bibinfo{author}{\bibfnamefont{J.~M.} \bibnamefont{Elzerman}},
  \bibinfo{author}{\bibfnamefont{R.~N.} \bibnamefont{Schouten}},
  \bibinfo{author}{\bibfnamefont{L.~H.~W.} \bibnamefont{van Beveren}},
  \bibinfo{author}{\bibfnamefont{R.}~\bibnamefont{Hanson}}, \bibnamefont{and}
  \bibinfo{author}{\bibfnamefont{L.~P.} \bibnamefont{Kouwenhoven}},
  \bibinfo{journal}{Appl. Phys. Lett.} \textbf{\bibinfo{volume}{85}},
  \bibinfo{pages}{4394} (\bibinfo{year}{2004}).

\bibitem[{\citenamefont{MacLean et~al.}(2007)\citenamefont{MacLean, Amasha,
  Radu, Zumb\"uhl, Kastner, Hanson, and Gossard}}]{MacLean2007}
\bibinfo{author}{\bibfnamefont{K.}~\bibnamefont{MacLean}},
  \bibinfo{author}{\bibfnamefont{S.}~\bibnamefont{Amasha}},
  \bibinfo{author}{\bibfnamefont{I.~P.} \bibnamefont{Radu}},
  \bibinfo{author}{\bibfnamefont{D.~M.} \bibnamefont{Zumb\"uhl}},
  \bibinfo{author}{\bibfnamefont{M.~A.} \bibnamefont{Kastner}},
  \bibinfo{author}{\bibfnamefont{M.~P.} \bibnamefont{Hanson}},
  \bibnamefont{and} \bibinfo{author}{\bibfnamefont{A.~C.}
  \bibnamefont{Gossard}}, \bibinfo{journal}{Phys. Rev. Lett.}
  \textbf{\bibinfo{volume}{98}}, \bibinfo{pages}{036802}
  (\bibinfo{year}{2007}).

\bibitem[{\citenamefont{Taubert et~al.}(2008)\citenamefont{Taubert,
  Pioro-Ladri\`ere, Schr\"oer, Harbusch, Sachrajda, and Ludwig}}]{Taubert2008}
\bibinfo{author}{\bibfnamefont{D.}~\bibnamefont{Taubert}},
  \bibinfo{author}{\bibfnamefont{M.}~\bibnamefont{Pioro-Ladri\`ere}},
  \bibinfo{author}{\bibfnamefont{D.}~\bibnamefont{Schr\"oer}},
  \bibinfo{author}{\bibfnamefont{D.}~\bibnamefont{Harbusch}},
  \bibinfo{author}{\bibfnamefont{A.~S.} \bibnamefont{Sachrajda}},
  \bibnamefont{and} \bibinfo{author}{\bibfnamefont{S.}~\bibnamefont{Ludwig}},
  \bibinfo{journal}{Phys. Rev. Lett.} \textbf{\bibinfo{volume}{100}},
  \bibinfo{pages}{176805} (\bibinfo{year}{2008}).

\bibitem[{\citenamefont{Kuhlmann et~al.}(2013)\citenamefont{Kuhlmann, Houel,
  Ludwig, Greuter, Reuter, Wieck, Poggio, and Warburton}}]{Kuhlmann2013}
\bibinfo{author}{\bibfnamefont{A.~V.} \bibnamefont{Kuhlmann}},
  \bibinfo{author}{\bibfnamefont{J.}~\bibnamefont{Houel}},
  \bibinfo{author}{\bibfnamefont{A.}~\bibnamefont{Ludwig}},
  \bibinfo{author}{\bibfnamefont{L.}~\bibnamefont{Greuter}},
  \bibinfo{author}{\bibfnamefont{D.}~\bibnamefont{Reuter}},
  \bibinfo{author}{\bibfnamefont{A.~D.} \bibnamefont{Wieck}},
  \bibinfo{author}{\bibfnamefont{M.}~\bibnamefont{Poggio}}, \bibnamefont{and}
  \bibinfo{author}{\bibfnamefont{R.~J.} \bibnamefont{Warburton}},
  \bibinfo{journal}{Nature Physics} \textbf{\bibinfo{volume}{9}},
  \bibinfo{pages}{570} (\bibinfo{year}{2013}).

\bibitem[{\citenamefont{Imamo\={g}lu et~al.}(1999)\citenamefont{Imamo\={g}lu,
  Awschalom, Burkard, DiVincenzo, Loss, Sherwin, and Small}}]{Imamoglu1999}
\bibinfo{author}{\bibfnamefont{A.}~\bibnamefont{Imamo\={g}lu}},
  \bibinfo{author}{\bibfnamefont{D.~D.} \bibnamefont{Awschalom}},
  \bibinfo{author}{\bibfnamefont{G.}~\bibnamefont{Burkard}},
  \bibinfo{author}{\bibfnamefont{D.~P.} \bibnamefont{DiVincenzo}},
  \bibinfo{author}{\bibfnamefont{D.}~\bibnamefont{Loss}},
  \bibinfo{author}{\bibfnamefont{M.}~\bibnamefont{Sherwin}}, \bibnamefont{and}
  \bibinfo{author}{\bibfnamefont{A.}~\bibnamefont{Small}},
  \bibinfo{journal}{Phys. Rev. Lett.} \textbf{\bibinfo{volume}{83}},
  \bibinfo{pages}{4204} (\bibinfo{year}{1999}).

\bibitem[{\citenamefont{Cooper et~al.}(2004)\citenamefont{Cooper, Steffen,
  McDermott, Simmonds, Oh, Hite, Pappas, and Martinis}}]{Cooper2004}
\bibinfo{author}{\bibfnamefont{K.~B.} \bibnamefont{Cooper}},
  \bibinfo{author}{\bibfnamefont{M.}~\bibnamefont{Steffen}},
  \bibinfo{author}{\bibfnamefont{R.}~\bibnamefont{McDermott}},
  \bibinfo{author}{\bibfnamefont{R.~W.} \bibnamefont{Simmonds}},
  \bibinfo{author}{\bibfnamefont{S.}~\bibnamefont{Oh}},
  \bibinfo{author}{\bibfnamefont{D.~A.} \bibnamefont{Hite}},
  \bibinfo{author}{\bibfnamefont{D.~P.} \bibnamefont{Pappas}},
  \bibnamefont{and} \bibinfo{author}{\bibfnamefont{J.~M.}
  \bibnamefont{Martinis}}, \bibinfo{journal}{Phys. Rev. Lett.}
  \textbf{\bibinfo{volume}{93}}, \bibinfo{pages}{180401}
  (\bibinfo{year}{2004}).

\bibitem[{\citenamefont{Siddiqi et~al.}(2006)\citenamefont{Siddiqi, Vijay,
  Metcalfe, Boaknin, Frunzio, Schoelkopf, and Devoret}}]{Siddiqi2006}
\bibinfo{author}{\bibfnamefont{I.}~\bibnamefont{Siddiqi}},
  \bibinfo{author}{\bibfnamefont{R.}~\bibnamefont{Vijay}},
  \bibinfo{author}{\bibfnamefont{M.}~\bibnamefont{Metcalfe}},
  \bibinfo{author}{\bibfnamefont{E.}~\bibnamefont{Boaknin}},
  \bibinfo{author}{\bibfnamefont{L.}~\bibnamefont{Frunzio}},
  \bibinfo{author}{\bibfnamefont{R.~J.} \bibnamefont{Schoelkopf}},
  \bibnamefont{and} \bibinfo{author}{\bibfnamefont{M.~H.}
  \bibnamefont{Devoret}}, \bibinfo{journal}{Phys. Rev. B}
  \textbf{\bibinfo{volume}{73}}, \bibinfo{pages}{054510}
  (\bibinfo{year}{2006}).

\bibitem[{\citenamefont{De et~al.}(2011)\citenamefont{De, Lang, Zhou, and
  Joynt}}]{De2011}
\bibinfo{author}{\bibfnamefont{A.}~\bibnamefont{De}},
  \bibinfo{author}{\bibfnamefont{A.}~\bibnamefont{Lang}},
  \bibinfo{author}{\bibfnamefont{D.}~\bibnamefont{Zhou}}, \bibnamefont{and}
  \bibinfo{author}{\bibfnamefont{R.}~\bibnamefont{Joynt}},
  \bibinfo{journal}{Phys. Rev. A} \textbf{\bibinfo{volume}{83}},
  \bibinfo{pages}{042331} (\bibinfo{year}{2011}).

\bibitem[{\citenamefont{De and Pryadko}(2013)}]{De2013prl}
\bibinfo{author}{\bibfnamefont{A.}~\bibnamefont{De}} \bibnamefont{and}
  \bibinfo{author}{\bibfnamefont{L.~P.} \bibnamefont{Pryadko}},
  \bibinfo{journal}{Phys. Rev. Lett.} \textbf{\bibinfo{volume}{110}},
  \bibinfo{pages}{070503} (\bibinfo{year}{2013}).

\bibitem[{\citenamefont{De and Pryadko}(2014)}]{De2014pra}
\bibinfo{author}{\bibfnamefont{A.}~\bibnamefont{De}} \bibnamefont{and}
  \bibinfo{author}{\bibfnamefont{L.~P.} \bibnamefont{Pryadko}},
  \bibinfo{journal}{Phys. Rev. A} \textbf{\bibinfo{volume}{89}},
  \bibinfo{pages}{032332} (\bibinfo{year}{2014}).

\bibitem[{\citenamefont{Majer et~al.}(2007)\citenamefont{Majer, Chow, Gambetta,
  Koch, Johnson, Schreier, Frunzio, Schuster, Houck, Wallraff
  et~al.}}]{Majer2007}
\bibinfo{author}{\bibfnamefont{J.}~\bibnamefont{Majer}},
  \bibinfo{author}{\bibfnamefont{J.~M.} \bibnamefont{Chow}},
  \bibinfo{author}{\bibfnamefont{J.~M.} \bibnamefont{Gambetta}},
  \bibinfo{author}{\bibfnamefont{J.}~\bibnamefont{Koch}},
  \bibinfo{author}{\bibfnamefont{B.~R.} \bibnamefont{Johnson}},
  \bibinfo{author}{\bibfnamefont{J.~A.} \bibnamefont{Schreier}},
  \bibinfo{author}{\bibfnamefont{L.}~\bibnamefont{Frunzio}},
  \bibinfo{author}{\bibfnamefont{D.~I.} \bibnamefont{Schuster}},
  \bibinfo{author}{\bibfnamefont{A.~A.} \bibnamefont{Houck}},
  \bibinfo{author}{\bibfnamefont{A.}~\bibnamefont{Wallraff}},
  \bibnamefont{et~al.}, \bibinfo{journal}{{ Nature}}
  \textbf{\bibinfo{volume}{449}}, \bibinfo{pages}{443} (\bibinfo{year}{2007}).

\bibitem[{\citenamefont{Fink et~al.}(2009)\citenamefont{Fink, Bianchetti, Baur,
  G\"oppl, Steffen, Filipp, Leek, Blais, and Wallraff}}]{Fink2009}
\bibinfo{author}{\bibfnamefont{J.~M.} \bibnamefont{Fink}},
  \bibinfo{author}{\bibfnamefont{R.}~\bibnamefont{Bianchetti}},
  \bibinfo{author}{\bibfnamefont{M.}~\bibnamefont{Baur}},
  \bibinfo{author}{\bibfnamefont{M.}~\bibnamefont{G\"oppl}},
  \bibinfo{author}{\bibfnamefont{L.}~\bibnamefont{Steffen}},
  \bibinfo{author}{\bibfnamefont{S.}~\bibnamefont{Filipp}},
  \bibinfo{author}{\bibfnamefont{P.~J.} \bibnamefont{Leek}},
  \bibinfo{author}{\bibfnamefont{A.}~\bibnamefont{Blais}}, \bibnamefont{and}
  \bibinfo{author}{\bibfnamefont{A.}~\bibnamefont{Wallraff}},
  \bibinfo{journal}{Phys. Rev. Lett.} \textbf{\bibinfo{volume}{103}},
  \bibinfo{pages}{083601} (\bibinfo{year}{2009}).

\bibitem[{\citenamefont{Chen et~al.}(2014)\citenamefont{Chen, Roushan, Sank,
  Neill, Lucero, Mariantoni, Barends, Chiaro, Kelly, Megrant
  et~al.}}]{Chen2014}
\bibinfo{author}{\bibfnamefont{Y.}~\bibnamefont{Chen}},
  \bibinfo{author}{\bibfnamefont{P.}~\bibnamefont{Roushan}},
  \bibinfo{author}{\bibfnamefont{D.}~\bibnamefont{Sank}},
  \bibinfo{author}{\bibfnamefont{C.}~\bibnamefont{Neill}},
  \bibinfo{author}{\bibfnamefont{E.}~\bibnamefont{Lucero}},
  \bibinfo{author}{\bibfnamefont{M.}~\bibnamefont{Mariantoni}},
  \bibinfo{author}{\bibfnamefont{R.}~\bibnamefont{Barends}},
  \bibinfo{author}{\bibfnamefont{B.}~\bibnamefont{Chiaro}},
  \bibinfo{author}{\bibfnamefont{J.}~\bibnamefont{Kelly}},
  \bibinfo{author}{\bibfnamefont{A.}~\bibnamefont{Megrant}},
  \bibnamefont{et~al.}, \bibinfo{journal}{arXiv:1403.6808}
  (\bibinfo{year}{2014}).

\bibitem[{\citenamefont{Shulman et~al.}(2012)\citenamefont{Shulman, Dial,
  Harvey, Bluhm, Umansky, and Yacoby}}]{Shulman2012}
\bibinfo{author}{\bibfnamefont{M.~D.} \bibnamefont{Shulman}},
  \bibinfo{author}{\bibfnamefont{O.~E.} \bibnamefont{Dial}},
  \bibinfo{author}{\bibfnamefont{S.~P.} \bibnamefont{Harvey}},
  \bibinfo{author}{\bibfnamefont{H.}~\bibnamefont{Bluhm}},
  \bibinfo{author}{\bibfnamefont{V.}~\bibnamefont{Umansky}}, \bibnamefont{and}
  \bibinfo{author}{\bibfnamefont{A.}~\bibnamefont{Yacoby}},
  \textbf{\bibinfo{volume}{336}}, \bibinfo{pages}{202} (\bibinfo{year}{2012}).

\bibitem[{\citenamefont{Cheng et~al.}(2008)\citenamefont{Cheng, Wang, and
  Joynt}}]{Cheng2008}
\bibinfo{author}{\bibfnamefont{B.}~\bibnamefont{Cheng}},
  \bibinfo{author}{\bibfnamefont{Q.-H.} \bibnamefont{Wang}}, \bibnamefont{and}
  \bibinfo{author}{\bibfnamefont{R.}~\bibnamefont{Joynt}},
  \bibinfo{journal}{Phys. Rev. A} \textbf{\bibinfo{volume}{78}},
  \bibinfo{pages}{022313} (\bibinfo{year}{2008}).

\bibitem[{\citenamefont{Joynt et~al.}(2011)\citenamefont{Joynt, Zhou, and
  Wang}}]{Joynt2011}
\bibinfo{author}{\bibfnamefont{R.}~\bibnamefont{Joynt}},
  \bibinfo{author}{\bibfnamefont{D.}~\bibnamefont{Zhou}}, \bibnamefont{and}
  \bibinfo{author}{\bibfnamefont{Q.-H.} \bibnamefont{Wang}},
  \bibinfo{journal}{International Journal of Modern Physics B}
  \textbf{\bibinfo{volume}{25}}, \bibinfo{pages}{2115} (\bibinfo{year}{2011}).

\bibitem[{\citenamefont{{Carmichael}}(1993)}]{Carmichael.book}
\bibinfo{author}{\bibfnamefont{H.}~\bibnamefont{{Carmichael}}},
  \bibinfo{journal}{An Open Systems Approach to Quantum Optics: Lectures
  Presented at the Universit{\'e} Libre de Bruxelles October 28 to November 4,
  1991, Lecture Notes in Physics Monographs, Volume 18.~ISBN
  978-3-540-56634-2.~Springer Berlin Heidelberg, 1993}
  \textbf{\bibinfo{volume}{18}} (\bibinfo{year}{1993}).

\bibitem[{\citenamefont{De and Joynt}(2013)}]{De2013pra}
\bibinfo{author}{\bibfnamefont{A.}~\bibnamefont{De}} \bibnamefont{and}
  \bibinfo{author}{\bibfnamefont{R.}~\bibnamefont{Joynt}},
  \bibinfo{journal}{Phys. Rev. A} \textbf{\bibinfo{volume}{87}},
  \bibinfo{pages}{042336} (\bibinfo{year}{2013}).

\bibitem[{\citenamefont{Kampen}(2007)}]{VanKampen.book}
\bibinfo{author}{\bibfnamefont{N.~G.~V.} \bibnamefont{Kampen}},
  \emph{\bibinfo{title}{Stochastic Processes in Physics and Chemistry}}
  (\bibinfo{publisher}{Elsevier}, \bibinfo{address}{Amsterdam},
  \bibinfo{year}{2007}), \bibinfo{edition}{3rd} ed.

\bibitem[{\citenamefont{Zhou and Joynt}(2010)}]{Zhou2010}
\bibinfo{author}{\bibfnamefont{D.}~\bibnamefont{Zhou}} \bibnamefont{and}
  \bibinfo{author}{\bibfnamefont{R.}~\bibnamefont{Joynt}},
  \bibinfo{journal}{Phys. Rev. A} \textbf{\bibinfo{volume}{81}},
  \bibinfo{pages}{010103} (\bibinfo{year}{2010}).

\bibitem[{\citenamefont{Tessier et~al.}(2003)\citenamefont{Tessier, Deutsch,
  Delgado, and Fuentes-Guridi}}]{Tessier2003}
\bibinfo{author}{\bibfnamefont{T.~E.} \bibnamefont{Tessier}},
  \bibinfo{author}{\bibfnamefont{I.~H.} \bibnamefont{Deutsch}},
  \bibinfo{author}{\bibfnamefont{A.}~\bibnamefont{Delgado}}, \bibnamefont{and}
  \bibinfo{author}{\bibfnamefont{I.}~\bibnamefont{Fuentes-Guridi}},
  \bibinfo{journal}{Phys. Rev. A} \textbf{\bibinfo{volume}{68}},
  \bibinfo{pages}{062316} (\bibinfo{year}{2003}).

\bibitem[{\citenamefont{Wootters}(1998)}]{Wootters1998}
\bibinfo{author}{\bibfnamefont{W.~K.} \bibnamefont{Wootters}},
  \bibinfo{journal}{Phys. Rev. Lett.} \textbf{\bibinfo{volume}{80}},
  \bibinfo{pages}{2245} (\bibinfo{year}{1998}).

\bibitem[{\citenamefont{Bylander et~al.}(2011)\citenamefont{Bylander,
  Gustavsson, Yan, Yoshihara, Harrabi, Fitch, Cory, Nakamura, Tsai, and
  Oliver}}]{Bylander2011}
\bibinfo{author}{\bibfnamefont{J.}~\bibnamefont{Bylander}},
  \bibinfo{author}{\bibfnamefont{S.}~\bibnamefont{Gustavsson}},
  \bibinfo{author}{\bibfnamefont{F.}~\bibnamefont{Yan}},
  \bibinfo{author}{\bibfnamefont{F.}~\bibnamefont{Yoshihara}},
  \bibinfo{author}{\bibfnamefont{K.}~\bibnamefont{Harrabi}},
  \bibinfo{author}{\bibfnamefont{G.}~\bibnamefont{Fitch}},
  \bibinfo{author}{\bibfnamefont{D.~G.} \bibnamefont{Cory}},
  \bibinfo{author}{\bibfnamefont{Y.}~\bibnamefont{Nakamura}},
  \bibinfo{author}{\bibfnamefont{J.-S.} \bibnamefont{Tsai}}, \bibnamefont{and}
  \bibinfo{author}{\bibfnamefont{W.~D.} \bibnamefont{Oliver}},
  \bibinfo{journal}{Nat. Phys.} \textbf{\bibinfo{volume}{7}},
  \bibinfo{pages}{565} (\bibinfo{year}{2011}).

\bibitem[{\citenamefont{Zhou and Joynt}(2012)}]{Zhou2012}
\bibinfo{author}{\bibfnamefont{D.}~\bibnamefont{Zhou}} \bibnamefont{and}
  \bibinfo{author}{\bibfnamefont{R.}~\bibnamefont{Joynt}},
  \bibinfo{journal}{Superconductor Science and Technology}
  \textbf{\bibinfo{volume}{25}}, \bibinfo{pages}{045003}
  (\bibinfo{year}{2012}).

\bibitem[{\citenamefont{Shalibo et~al.}(2010)\citenamefont{Shalibo, Rofe, Shwa,
  Zeides, Neeley, Martinis, and Katz}}]{Yoni2010}
\bibinfo{author}{\bibfnamefont{Y.}~\bibnamefont{Shalibo}},
  \bibinfo{author}{\bibfnamefont{Y.}~\bibnamefont{Rofe}},
  \bibinfo{author}{\bibfnamefont{D.}~\bibnamefont{Shwa}},
  \bibinfo{author}{\bibfnamefont{F.}~\bibnamefont{Zeides}},
  \bibinfo{author}{\bibfnamefont{M.}~\bibnamefont{Neeley}},
  \bibinfo{author}{\bibfnamefont{J.~M.} \bibnamefont{Martinis}},
  \bibnamefont{and} \bibinfo{author}{\bibfnamefont{N.}~\bibnamefont{Katz}},
  \bibinfo{journal}{Phys. Rev. Lett.} \textbf{\bibinfo{volume}{105}},
  \bibinfo{pages}{177001} (\bibinfo{year}{2010}).

\bibitem[{\citenamefont{{Geller} et~al.}(2014)\citenamefont{{Geller}, {Donate},
  {Chen}, {Neill}, {Roushan}, and {Martinis}}}]{Geller2014}
\bibinfo{author}{\bibfnamefont{M.~R.} \bibnamefont{{Geller}}},
  \bibinfo{author}{\bibfnamefont{E.}~\bibnamefont{{Donate}}},
  \bibinfo{author}{\bibfnamefont{Y.}~\bibnamefont{{Chen}}},
  \bibinfo{author}{\bibfnamefont{C.}~\bibnamefont{{Neill}}},
  \bibinfo{author}{\bibfnamefont{P.}~\bibnamefont{{Roushan}}},
  \bibnamefont{and} \bibinfo{author}{\bibfnamefont{J.~M.}
  \bibnamefont{{Martinis}}}, \bibinfo{journal}{ArXiv e-prints}
  (\bibinfo{year}{2014}), \eprint{1405.1915}.

\bibitem[{\citenamefont{Chiorescu et~al.}(2004)\citenamefont{Chiorescu, Bertet,
  Semba, Nakamura, Harmans, and Mooij}}]{Chiorescu2004}
\bibinfo{author}{\bibfnamefont{I.}~\bibnamefont{Chiorescu}},
  \bibinfo{author}{\bibfnamefont{P.}~\bibnamefont{Bertet}},
  \bibinfo{author}{\bibfnamefont{K.}~\bibnamefont{Semba}},
  \bibinfo{author}{\bibfnamefont{Y.}~\bibnamefont{Nakamura}},
  \bibinfo{author}{\bibfnamefont{C.~J. P.~M.} \bibnamefont{Harmans}},
  \bibnamefont{and} \bibinfo{author}{\bibfnamefont{J.~E.} \bibnamefont{Mooij}},
  \bibinfo{journal}{Nature} \textbf{\bibinfo{volume}{431}},
  \bibinfo{pages}{159} (\bibinfo{year}{2004}).

\bibitem[{\citenamefont{Bourassa et~al.}(2009)\citenamefont{Bourassa, Gambetta,
  Abdumalikov, Astafiev, Nakamura, and Blais}}]{Bourassa2009}
\bibinfo{author}{\bibfnamefont{J.}~\bibnamefont{Bourassa}},
  \bibinfo{author}{\bibfnamefont{J.~M.} \bibnamefont{Gambetta}},
  \bibinfo{author}{\bibfnamefont{A.~A.} \bibnamefont{Abdumalikov}},
  \bibinfo{author}{\bibfnamefont{O.}~\bibnamefont{Astafiev}},
  \bibinfo{author}{\bibfnamefont{Y.}~\bibnamefont{Nakamura}}, \bibnamefont{and}
  \bibinfo{author}{\bibfnamefont{A.}~\bibnamefont{Blais}},
  \bibinfo{journal}{Phys. Rev. A} \textbf{\bibinfo{volume}{80}},
  \bibinfo{pages}{032109} (\bibinfo{year}{2009}).

\bibitem[{\citenamefont{Fedorov et~al.}(2010)\citenamefont{Fedorov, Feofanov,
  Macha, Forn-Diaz, Harmans, and Mooij}}]{Fedorov2010}
\bibinfo{author}{\bibfnamefont{A.}~\bibnamefont{Fedorov}},
  \bibinfo{author}{\bibfnamefont{A.~K.} \bibnamefont{Feofanov}},
  \bibinfo{author}{\bibfnamefont{P.}~\bibnamefont{Macha}},
  \bibinfo{author}{\bibfnamefont{P.}~\bibnamefont{Forn-Diaz}},
  \bibinfo{author}{\bibfnamefont{C.~J. P.~M.} \bibnamefont{Harmans}},
  \bibnamefont{and} \bibinfo{author}{\bibfnamefont{J.~E.} \bibnamefont{Mooij}},
  \bibinfo{journal}{Phys. Rev. Lett.} \textbf{\bibinfo{volume}{105}},
  \bibinfo{pages}{060503} (\bibinfo{year}{2010}).

\bibitem[{\citenamefont{Chiorescu et~al.}(2003)\citenamefont{Chiorescu,
  Nakamura, Harmans, and Mooij}}]{Chiorescu2003}
\bibinfo{author}{\bibfnamefont{I.}~\bibnamefont{Chiorescu}},
  \bibinfo{author}{\bibfnamefont{Y.}~\bibnamefont{Nakamura}},
  \bibinfo{author}{\bibfnamefont{C.~J. P.~M.} \bibnamefont{Harmans}},
  \bibnamefont{and} \bibinfo{author}{\bibfnamefont{J.~E.} \bibnamefont{Mooij}},
  \bibinfo{journal}{Science} \textbf{\bibinfo{volume}{299}},
  \bibinfo{pages}{1869} (\bibinfo{year}{2003}), ISSN \bibinfo{issn}{1095-9203}.

\bibitem[{\citenamefont{Majer et~al.}(2005)\citenamefont{Majer, Paauw, ter
  Haar, Harmans, and Mooij}}]{Majer2005}
\bibinfo{author}{\bibfnamefont{J.~B.} \bibnamefont{Majer}},
  \bibinfo{author}{\bibfnamefont{F.~G.} \bibnamefont{Paauw}},
  \bibinfo{author}{\bibfnamefont{A.~C.~J.} \bibnamefont{ter Haar}},
  \bibinfo{author}{\bibfnamefont{C.~J. P.~M.} \bibnamefont{Harmans}},
  \bibnamefont{and} \bibinfo{author}{\bibfnamefont{J.~E.} \bibnamefont{Mooij}},
  \bibinfo{journal}{Phys. Rev. Lett.} \textbf{\bibinfo{volume}{94}},
  \bibinfo{pages}{090501} (\bibinfo{year}{2005}).

\bibitem[{\citenamefont{van~der Ploeg et~al.}(2007)\citenamefont{van~der Ploeg,
  Izmalkov, van~den Brink, H\"ubner, Grajcar, Il'ichev, Meyer, and
  Zagoskin}}]{vanderPloeg2007}
\bibinfo{author}{\bibfnamefont{S.~H.~W.} \bibnamefont{van~der Ploeg}},
  \bibinfo{author}{\bibfnamefont{A.}~\bibnamefont{Izmalkov}},
  \bibinfo{author}{\bibfnamefont{A.~M.} \bibnamefont{van~den Brink}},
  \bibinfo{author}{\bibfnamefont{U.}~\bibnamefont{H\"ubner}},
  \bibinfo{author}{\bibfnamefont{M.}~\bibnamefont{Grajcar}},
  \bibinfo{author}{\bibfnamefont{E.}~\bibnamefont{Il'ichev}},
  \bibinfo{author}{\bibfnamefont{H.-G.} \bibnamefont{Meyer}}, \bibnamefont{and}
  \bibinfo{author}{\bibfnamefont{A.~M.} \bibnamefont{Zagoskin}},
  \bibinfo{journal}{Phys. Rev. Lett.} \textbf{\bibinfo{volume}{98}},
  \bibinfo{pages}{057004} (\bibinfo{year}{2007}).

\bibitem[{\citenamefont{Makhlin et~al.}(2001)\citenamefont{Makhlin, Sch\"on,
  and Shnirman}}]{Makhlin2001}
\bibinfo{author}{\bibfnamefont{Y.}~\bibnamefont{Makhlin}},
  \bibinfo{author}{\bibfnamefont{G.}~\bibnamefont{Sch\"on}}, \bibnamefont{and}
  \bibinfo{author}{\bibfnamefont{A.}~\bibnamefont{Shnirman}},
  \bibinfo{journal}{Rev. Mod. Phys.} \textbf{\bibinfo{volume}{73}},
  \bibinfo{pages}{357} (\bibinfo{year}{2001}).

\bibitem[{\citenamefont{Wendin and Shumeiko}(2007)}]{Wendin2007}
\bibinfo{author}{\bibfnamefont{G.}~\bibnamefont{Wendin}} \bibnamefont{and}
  \bibinfo{author}{\bibfnamefont{V.~S.} \bibnamefont{Shumeiko}},
  \bibinfo{journal}{Low Temperature Physics} \textbf{\bibinfo{volume}{33}},
  \bibinfo{pages}{724 } (\bibinfo{year}{2007}).

\bibitem[{\citenamefont{Byron and Fuller}(1992)}]{ByronFuller.book}
\bibinfo{author}{\bibfnamefont{F.}~\bibnamefont{Byron}} \bibnamefont{and}
  \bibinfo{author}{\bibfnamefont{R.}~\bibnamefont{Fuller}},
  \emph{\bibinfo{title}{Mathematics of Classical and Quantum Physics}}, Dover
  books on physics and chemistry (\bibinfo{publisher}{Dover Publications},
  \bibinfo{year}{1992}), ISBN \bibinfo{isbn}{9780486671642}.

\bibitem[{\citenamefont{Blum}(2010)}]{Blum.book}
\bibinfo{author}{\bibfnamefont{K.}~\bibnamefont{Blum}},
  \emph{\bibinfo{title}{Density Matrix Theory and Applications}}, Physics of
  Atoms and Molecules (\bibinfo{publisher}{Springer}, \bibinfo{year}{2010}),
  ISBN \bibinfo{isbn}{9781441932570}.

\bibitem[{\citenamefont{Xiang et~al.}(2013)\citenamefont{Xiang, Ashhab, You,
  and Nori}}]{Xiang2013RMP}
\bibinfo{author}{\bibfnamefont{Z.-L.} \bibnamefont{Xiang}},
  \bibinfo{author}{\bibfnamefont{S.}~\bibnamefont{Ashhab}},
  \bibinfo{author}{\bibfnamefont{J.~Q.} \bibnamefont{You}}, \bibnamefont{and}
  \bibinfo{author}{\bibfnamefont{F.}~\bibnamefont{Nori}},
  \bibinfo{journal}{Rev. Mod. Phys.} \textbf{\bibinfo{volume}{85}},
  \bibinfo{pages}{623} (\bibinfo{year}{2013}).

\bibitem[{\citenamefont{Paik et~al.}(2011)\citenamefont{Paik, Schuster, Bishop,
  Kirchmair, Catelani, Sears, Johnson, Reagor, Frunzio, Glazman
  et~al.}}]{Paik2011}
\bibinfo{author}{\bibfnamefont{H.}~\bibnamefont{Paik}},
  \bibinfo{author}{\bibfnamefont{D.~I.} \bibnamefont{Schuster}},
  \bibinfo{author}{\bibfnamefont{L.~S.} \bibnamefont{Bishop}},
  \bibinfo{author}{\bibfnamefont{G.}~\bibnamefont{Kirchmair}},
  \bibinfo{author}{\bibfnamefont{G.}~\bibnamefont{Catelani}},
  \bibinfo{author}{\bibfnamefont{A.~P.} \bibnamefont{Sears}},
  \bibinfo{author}{\bibfnamefont{B.~R.} \bibnamefont{Johnson}},
  \bibinfo{author}{\bibfnamefont{M.~J.} \bibnamefont{Reagor}},
  \bibinfo{author}{\bibfnamefont{L.}~\bibnamefont{Frunzio}},
  \bibinfo{author}{\bibfnamefont{L.~I.} \bibnamefont{Glazman}},
  \bibnamefont{et~al.}, \bibinfo{journal}{Phys. Rev. Lett.}
  \textbf{\bibinfo{volume}{107}}, \bibinfo{pages}{240501}
  (\bibinfo{year}{2011}).

\bibitem[{\citenamefont{Stern et~al.}(2014)\citenamefont{Stern, Kubo, Grezes,
  Bienfait, Vion, Esteve, and Bertet}}]{Stern2014}
\bibinfo{author}{\bibfnamefont{M.}~\bibnamefont{Stern}},
  \bibinfo{author}{\bibfnamefont{Y.}~\bibnamefont{Kubo}},
  \bibinfo{author}{\bibfnamefont{C.}~\bibnamefont{Grezes}},
  \bibinfo{author}{\bibfnamefont{A.}~\bibnamefont{Bienfait}},
  \bibinfo{author}{\bibfnamefont{D.}~\bibnamefont{Vion}},
  \bibinfo{author}{\bibfnamefont{D.}~\bibnamefont{Esteve}}, \bibnamefont{and}
  \bibinfo{author}{\bibfnamefont{P.}~\bibnamefont{Bertet}},
  \bibinfo{journal}{arXiv preprint arXiv:1403.3871}  (\bibinfo{year}{2014}).

\end{thebibliography}

\end{document}